\documentclass[journal,twocolumn, 10pt]{IEEEtran}

\usepackage{soul,xcolor}
\usepackage{bm}
\include{graphics}
\include{amsmath}
\usepackage[normalem]{ulem}
\usepackage{multirow,enumitem}
\usepackage{algorithm}
\usepackage{algpseudocode}
\usepackage{verbatim}
\usepackage[latin1,utf8]{inputenc}
\usepackage{amsmath}
\usepackage{amsfonts}
\usepackage{amssymb}
\usepackage{mathrsfs}
\usepackage{booktabs}
\usepackage{url}
\usepackage{ifthen}
\usepackage{xspace}
\usepackage{dsfont}
\usepackage{bbm}
\usepackage{cite}
\usepackage{epsfig}
\usepackage{epstopdf}
\usepackage{array}
\usepackage{algpseudocode}
\usepackage{algorithm}
\usepackage{breqn}
\usepackage{esint}
\usepackage{amsthm}
\usepackage{setspace}
\usepackage{lipsum}

\theoremstyle{plain}

\theoremstyle{definition}

\theoremstyle{remark}

\usepackage{footnote}
\theoremstyle{plain}
\newtheorem{assumption}{Assumption}

\newcommand{\round}[1]{\ensuremath{\left\lfloor#1\right\rceil}}
\usepackage{bbold}
\usepackage{todonotes}
\newtheorem{definition}{Definition}
\usepackage{stmaryrd}
\usepackage{algorithm,algcompatible}
\algnewcommand\INPUT{\item[\textbf{Input:}]}%
\algnewcommand\OUTPUT{\item[\textbf{Output:}]}%
\usepackage{lipsum}
\usepackage{mathtools}
\usepackage{cuted}
\usepackage{dblfloatfix}
\usepackage{subfigure}

\usepackage{graphicx}
\usepackage{epstopdf}
\newcommand{\mathleft}{\@fleqntrue\@mathmargin0pt}
\newcommand{\mathcenter}{\@fleqnfalse}

\ifCLASSINFOpdf

\else

\fi

\begin{document}
	\setulcolor{red}
\setul{red}{2pt}
\setstcolor{red}

	\title{
	Fast Position-Aided MIMO Beam Training\\ via Noisy Tensor Completion
	}
	\author{Tzu-Hsuan~{Chou}, Nicol\`{o} {Michelusi}, David J. {Love}, and James V. {Krogmeier}
		\thanks{This work was supported in part by the National Science Foundation under grant CNS-1642982 and CCF-1816013.}		
		\thanks{A preliminary version of this paper was presented at the IEEE International	Conference on Communications 2020 \cite{ICC20}.}
		\thanks{The authors are with the School of Electrical and Computer Engineering, Purdue University, West Lafayette, IN, USA; emails: \{chou59, michelus, djlove, jvk\}@purdue.edu.}%
		\vspace{-5mm}
	}

	\maketitle
	
	\begin{abstract}
	In this paper, a data-driven position-aided approach is
	proposed to reduce the training overhead in MIMO systems, by leveraging side information and on-the-field measurements.
	A data tensor is constructed by collecting beam-training measurements on a subset of positions and beams, and a hybrid noisy tensor completion (HNTC) algorithm is proposed to predict the received power across the coverage area, which exploits both the spatial smoothness and the low-rank property of MIMO channels.
	A recommendation algorithm based on the completed tensor, beam subset selection (BSS), is proposed to achieve fast and accurate beam-training.
	Besides, a grouping-based BSS algorithm is proposed to combat the detrimental effect of noisy positional information. 
	Numerical results evaluated with the Quadriga channel simulator at 60 GHz millimeter-wave channels show that the proposed BSS recommendation algorithm in combination with HNTC achieve accurate received power predictions, enabling beam-alignment with small overhead:
	given power measurements on 40\% of possible discretized positions, HNTC-based BSS attains a probability of correct alignment of 91\%, with only 2\% of trained beams, as opposed to a state-of-the-art position-aided beam-alignment scheme which achieves 54\% correct alignment in the same configuration.
	Finally, an online HNTC method via warm-start is proposed, that alleviates the computational complexity by 50\%, with no degradation in prediction accuracy.
	
	\end{abstract}
	
	
\vspace{-3mm}
	\section{Introduction}
	Future wireless networks will be required to sustain high data rates, low latencies, and improved power efficiencies for users (UEs) in a wide variety of conditions \cite{andrews2014will}.
	In the last few years,
	 millimeter-wave (mmWave) communication has been gaining more attention as a viable solution to address the throughput enhancement challenges faced by 5G and beyond wireless networks, thanks to the large bandwidth availability \cite{rappaport2013millimeter,heath2016overview,boccardi2014five}. 
	Nonetheless, mmWave communication suffers from severe pathloss due to the high frequency of the signal, hence it requires the use of narrow beam communication to compensate the signal attenuation. This is achieved with the use of large antenna arrays at the transmitter and receiver via massive multiple-input and multiple-output (MIMO) systems.	However, traditional massive MIMO channel estimation methods are impractical, especially in mobile environments, because of the unacceptably large overhead induced by the large number of antennas.

	The overhead of massive MIMO channel estimation, either operating at mmWave or sub-6GHz frequencies, can be reduced by exploiting the spatial sparsity of the channel, resulting from few dominant clusters or paths in the angular domain and the high spatial resolution \cite{xie2016overview}.
	This sparsity allows to predesign a set of directional beams pointing in specific directions, so that beam training can be done directly on this predefined set, rather than on the MIMO channel matrix.
	The standard approach to beam training is to search in an exhaustive fashion through all possible combinations of transmit and receive beams, to determine the beam pair with maximum signal power. However, this approach incurs enormous overhead, due to the large number of possible beam combinations that need to be periodically trained.
	To reduce the overhead of exhaustive search, several beam-alignment schemes have been proposed in the literature, ranging from \textit{feedback-based methods} \cite{6600706,hussain2018energy,9013578}, \textit{AOAs/AODs estimation} \cite{alkhateeb2015compressed,alkhateeb2014channel,marzi2016compressive}, to \textit{data-assisted schemes} \cite{gonzalez2016radar,8642397,8101513,8198818,Va2018,8734054,8823977}.
	\textit{Feedback-based methods} adapt the beam-training procedure in an online fashion based on feedback collected.
	\textit{AOAs/AODs estimation} leverages the spatial sparsity of the mmWave channel via compressed sensing techniques to recover AOAs/AODs and gains of the channel paths.
	
	In a \textit{data-assisted scheme}, the beam-training process is aided by side information from the available sensors other than mmWave communication, such as radar \cite{gonzalez2016radar}, LIDAR \cite{8642397}, lower-frequency communication \cite{8101513,8198818}, or GPS position information \cite{Va2018,8734054,8823977}.
	Our work uses the positional information.
	Inverse fingerprinting for beam alignment is proposed in \cite{Va2018}, using prior measurements at a given position to provide a set of candidate beam directions at the same position.
	Their work demonstrated that the positional information of the user can be utilized to predict a small set of candidate beams to train, based on long-term channel information collected in the database.
	The mmWave beam selection problem at specific locations is formulated in \cite{8734054} as a machine learning classification problem using past beam training measurements with the situational awareness, which captures the environmental condition by encoding the obstacles' coordinates.
	In \cite{8823977}, a multiple-fingerprint beam alignment method which intelligently selects the fingerprint beam alignment based on the traffic density via learning is proposed.
	
	However, all works on positional data-assisted beam-alignment techniques \cite{Va2018,8734054,8823977} fail to provide the channel information in the positions whose prior measurements are not available (new positions).
	For this reason, the existing approaches require the collection of an extremely large amount of channel measurements to cover a given service region, which may not be practical.	
	Furthermore, the user's position acquired by the process of global navigation satellite system (GNSS)/global positioning system (GPS) estimation is possibly noisy due to the estimation error and the user's mobility \cite{roy2014smartphone,maschietti2017robust}.
	Positional estimation error degrades the performance of position-aided data-assisted schemes.
	To address these general problems, we seek to do the channel prediction in new positions based on the channel information in observed positions, and propose a beam-recommendation scheme robust against errors in positional information.
	
	Recently, channel cartography (channel charting) \cite{7956220,8444621} of the wireless network has been investigated.
	The prior channel measurements on positions provide helpful information for reducing the training overhead of the UE in the service area.
	The work \cite{7956220} estimates the spatial loss field map from the measurements on few positions aided by a matrix completion technique which leverages the low-rank structure and sparsity.
	The channel gain between two points is modeled as the tomographic accumulation of the spatial loss field along the propagation path.
	The work \cite{8444621} proposed a channel charting framework for locating users in multiple-antenna wireless networks, learning the relationship between channel state information and user's location with the tools from dimensionality reduction, manifold learning, and neural networks.
	In our work, we borrow ideas to learn the radio geometry by constructing a data tensor recording the average received power of beams on few positions.
	The average received power tensor is recovered by the proposed tensor completion, \emph{hybrid noisy tensor completion} (HNTC), exploiting the low-rank structure of the channel sparsity and the smoothness from the spatial correlation.
	
	\vspace{-1em}
	\subsection{Related Work}	
	Real world data often exhibit various structural properties, which enable reconstruction from sparse samples \cite{6138863,cai2010singular}.
	In the two-dimensional (2D) data structure (matrix), rank is a powerful factor capturing the global information.
	Low-rank matrix approximation has been intensely investigated \cite{cai2010singular,recht2010guaranteed,candes2010power,10.1145/2184319.2184343,5454406}, and various algorithmic approaches capable of estimating missing values have been developed.	
	In \cite{cai2010singular,recht2010guaranteed}, the nuclear norm has been shown as the tightest convex envelope for the matrix rank function; and an efficient algorithm, singular value thresholding (SVT), was proposed.
	In \cite{10.1145/2184319.2184343,candes2010power}, the authors showed that the rank minimization problem can be solved by minimizing the nuclear norm under certain conditions, which justified the validity of using nuclear norm as the surrogate of the rank function theoretically.
	Low-rank matrix completion with noisy data is investigated in \cite{5454406}.
	Smoothness \cite{dai2009physics, han2014linear} is also an important feature aiding the matrix completion.
	The work \cite{dai2009physics} considered  total variation as the objective function for the matrix completion problem. 
	In \cite{han2014linear}, the authors exploited both the low-rank and smoothness simultaneously for the matrix completion.
	In addition to the matrix approximation, there are works developed for tensors, which are higher dimensional extensions of matrices.
	Tensor completion has gained attention recently due to its multidimensional character in describing intricate datasets, with many applications.
	Most works developed the tensor completion approaches for image processing because the image color is composed of red/green/blue, which makes tensor a suitable structure.
	Recently, various models have been used for tensor completion including Tucker rank \cite{tomioka2010estimation,gandy2011tensor,6138863,8625383,liu2019low}, canonical decomposition/parallel factors (CANDECOMP/PARAFAC) rank (also known as CP rank) \cite{liu2019low,yokota2016smooth}, and total variation (TV) \cite{8625383,yokota2016smooth,li2017low}.
	TV tensor completion takes advantage of the data smoothness, and the remaining models exploit the low-rank of the tensor.
	However, computing the rank of the tensor is an NP hard problem \cite{hillar2013most}, so the best model for low-rank tensor approximation generally does not exist. 
	A suitable solution for tensor completion depends on the application.
	
	Tensor completion algorithms using both low-rank and smoothness are investigated in \cite{yokota2016smooth,8625383}.
	The work \cite{8625383} formulated the problem considering the low Tucker rank and data smoothness of the tensor model; and an algorithm, LRTV-PDS, was proposed using the primal-dual splitting approach.
	In \cite{yokota2016smooth}, the authors proposed an algorithm, SPCTV, based on the low CP rank with the smoothness constraint.
	The works \cite{8625383,yokota2016smooth} considered the low-rank property and the smoothness consistent on all dimensions. 
	However, the properties of the tensor dimensions are based on the assumed structural properties.
	It is possible that the low-rank property only exists in some dimensions, and the data of the remaining dimensions are smooth.
	Moreover, the tensor completion problem with noisy data is investigated in \cite{8625383}.
	In this work, we seek to devise a noisy tensor completion algorithm which exploits the low-rank and smoothness properties on separate dimensions for noisy measurement data.
	

	\vspace{-1em}
	\subsection{Contribution}
	We develop a new tensor completion algorithm, HNTC, to enable fast and accurate MIMO beam training.
	To chart the MIMO channel conditions in the service area, we construct a data tensor by collecting the received power on a subset of positions and beams.
	The data tensor has the dimensions corresponding to the positional and beam information, respectively.
	Using the sparsity of MIMO channels \cite{rappaport2013millimeter,heath2016overview,xie2016overview}, the data tensor exhibits the low-rank property in the beam dimensions.
	On the other hand, the spatial correlation of the channel induces smoothness across the position dimension of the data tensor.
	To account for noise in the received signal and random fluctuations in the channel, we formulate the data recovery as a noisy tensor completion problem considering locally low-rank and spatial smoothness in distinct dimensions, which is a convex optimization problem with a noise inequality constraint.	
	
	
	The contributions of the paper are detailed as follow:
	\begin{itemize}
		\item We propose a tensor completion algorithm, HNTC, that uses the low-rank and smoothness properties across the different dimensions in the data to reconstruct channel properties from noisy measurements.
		
		\item We develop an online version of HNTC using warm-start.
		
		\item We propose a position-aided beam recommendation algorithm for fast and accurate MIMO beam-training, beam subset selection (BSS), which uses the  tensor completed via HNTC to recommend a small subset of beams to train.
		
		\item We propose a grouping-based BSS (G-BSS) algorithm to combat the effect of errors in positional information.
	\end{itemize}

	The rest of the paper is organized as follows.
	Section \ref{sec_prelim_tensor} reviews the preliminaries on tensor completion.
	Section \ref{sec_motivation_ex} introduces a data-driven approach for learning the channel conditions, which motivates a new noisy tensor completion problem.
	Section \ref{TC_problem} proposes the algorithm HNTC aided by alternating direction method of multipliers (ADMM) \cite{boyd2011distributed}, and Section \ref{Online_TC_algorithm} introduces online HNTC.
	Section \ref{complexity} discusses the computational complexity.
	Section \ref{sec_numerical_result} evaluates the performance of HNTC and position-aided beam-alignment.
	Section \ref{sec_conclusion} concludes the paper.

	\section{Preliminaries on Tensor Completion}
	\label{sec_prelim_tensor}
	Tensors are commonly used in many areas of engineering and science \cite{kolda2009tensor}, but they have received only limited interest in communication theory.
	In the following, we introduce the tensor notation and terminology used throughout the rest of the paper.
		\vspace{-5mm}
	\subsection{Notations}\label{tensor_intro}
	Bold uppercase letters $\mathbf X$ denote matrices, and calligraphic letters $\mathcal X$ represent tensors.
	$\mathbf{X}\otimes \mathbf{Y}$ is the Kronecker product of $\mathbf{X}$ and $\mathbf{Y}$.	
	An $M$-th order tensor is defined as $\mathcal{X}\in\mathbb{R}^{I_1 \times I_2\times\dots\times I_M}$, with $M$ being the number of dimensions (a matrix can be interpreted as a second order tensor).
	Given an $M$-th order tensor $\mathcal{X}$,	we denote its $\mathbf i_{1:M}=(i_1,i_2,\dots,i_M)$-th element as $\mathcal{X}(\mathbf i_{1:M})=\mathcal{X}(i_1,i_2,\dots,i_M)$.
	The inner product of two tensors $\mathcal{X},\mathcal{Y}\in\mathbb{R}^{I_1 \times\dots\times I_M}$ is defined as
	\begin{equation*}
		\left<\mathcal{X},\mathcal{Y}\right>\triangleq
		\sum_{\mathbf i_{1:M}}\mathcal X(\mathbf i_{1:M})\mathcal Y(\mathbf i_{1:M})
	\end{equation*}	
	and the Frobenius norm of $\mathcal X$ is denoted by
	$\lVert\mathcal{X}\rVert_F\triangleq\sqrt{\left<\mathcal{X},\mathcal{X}\right>}.$
	The mode-$m$ unfolding (or the mode-$m$ matricization) of a tensor 
	$\mathcal{X}$ is the matrix $\mathcal{X}_{(m)}\in\mathbb{R}^{I_m\times{J_{M+1}}}$ with entries $\mathcal{X}_{(m)}\left(i_m,j\right)=\mathcal{X}(\mathbf i_{1:M}),\ \forall \mathbf i_{1:M}$, where $$j=1+\sum_{k=1,k\neq m}^{M}(i_k-1)J_k,$$
	with $J_k=\prod_{d=1,d\neq m}^{k-1}I_d$.
	The inverse operation of folding recovers the original tensor from its unfolded representation, and is denoted as $\text{fold}_m(\mathcal{X}_{(m)}) \triangleq \mathcal{X}$.
		\vspace{-5mm}
	\subsection{Preliminaries} \label{sec_prelim}
	The low-rank tensor completion is a natural extension of low-rank matrix completion to more than two dimensions.
	Given an $M$-th order incomplete tensor $\mathcal{T}$ with $\Psi$ as the 
	indices of known elements, the low-rank tensor completion problem is formulated as
	\begin{align}
	&\min_{\mathcal{X}}\ \mathrm{rank}(\mathcal{X})\ \mathrm{s.t.}\ 
	\mathcal{X}_\Psi = \mathcal{T}_\Psi,
	\end{align}
	where $\mathrm{rank}(\mathcal{X})$ is the smallest number of rank-one tensors needed to sum together to construct $\mathcal{X}$ (also known as CP rank), which is an analogue to the definition of matrix rank \cite{kolda2009tensor}.
	Here, a tensor  $\mathcal{X}\in\mathbb{R}^{I_1\times\cdots\times I_M}$ is rank-one if it can be expressed as the outer product of $M$ vectors, $\mathbf{a}_d\in\mathbb{R}^{I_d\times 1},\ d=1,\dots, M$, so that $\mathcal{X}(i_1,i_2,\cdots,i_M)=\prod_{d=1}^M\mathbf{a}_d(i_d)$ \cite{kolda2009tensor}.

	However, finding the rank of a specific tensor is an NP-hard problem \cite{hillar2013most}, and the best rank approximation may not exist \cite{kolda2001orthogonal}.
	For this reason, alternative definitions of rank have been used in the literature.
	A well developed definition is the Tucker rank \cite{kolda2009tensor}, defined as the rank of the unfolding matrices, $\text{rank}(\mathcal{X}_{(n)})$, $n=1,\dots,M$.
	The work \cite{6138863} defined the tensor nuclear norm as a weighted sum of the Tucker ranks of all the unfolding matrices:
	\begin{equation} \label{def_tensor_nuclear_norm}
	\lVert\mathcal{X}\rVert_*=\sum_{k=1}^{M} \alpha_k\lVert\mathcal{X}_{(k)}\rVert_*,
	\end{equation}	
	where the constant coefficients $\alpha_k>0$ and $\sum_{k=1}^M \alpha_k =1$; and then proposed the low-rank tensor completion problem as
	\begin{align} \label{opt_prob_Liu}
	&\min_{\mathcal{X}}\ \sum_{k=1}^{M} \alpha_k\lVert\mathcal{X}_{(k)}\rVert_*\ 
	\mathrm{s.t.}\ 
	\mathcal{X}_\Psi = \mathcal{T}_\Psi.  
	\end{align}
	The high accuracy low-rank tensor completion (HaLRTC) 
	method \cite{6138863} was proposed to solve \eqref{opt_prob_Liu} via ADMM.

	Smoothness has been considered a useful property to aid matrix completion \cite{han2014linear}, especially with high ratio of missing elements.
	In \cite{han2014linear}, the authors proposed a combination of low-rank and smoothness minimization for the matrix completion, termed the \textit{linear total variation approximate regularized nuclear norm minimization} problem (LTVNN).
	Given an incomplete matrix $\mathbf M$ with $\Omega$ as the indices of known elements, the completion problem is formulated as
	\begin{equation}\label{MC_LTV}
		\min_{\mathbf{X}}\ \lVert\mathbf{X}\rVert_*+\gamma\lVert\mathbf{X}\rVert_{LTV},\ \mathrm{s.t.}\ \mathbf{X}_{\Omega}=\mathbf{M}_{\Omega},
	\end{equation}
	where $\gamma$ is the trade-off factor between low-rank and smoothness, and $\lVert\mathbf{X}\rVert_{LTV}$ is the linear total variation, defined as 
	\begin{equation}
		\lVert\mathbf{X}\rVert_{LTV}=\sum_{i,j}{\delta_1^2(\mathbf{X}(i,j))+\delta_2^2(\mathbf{X}(i,j))},
	\end{equation}
	where $\delta_1(\mathbf{X}(i,j))\equiv\mathbf{X}(i+1,j)-\mathbf{X}(i,j)$ and 
	$\delta_2(\mathbf{X}(i,j))\equiv\mathbf{X}(i,j+1)-\mathbf{X}(i,j)$.
	This optimization problem \eqref{MC_LTV} can be solved with the ADMM method \cite{boyd2011distributed}.


	For the tensor completion, there were previous works \cite{8625383,yokota2016smooth} considering both low-rank and smoothness minimization.
	The work \cite{8625383} proposed the algorithm considering low Tucker rank model with data smoothness, called LRTV-PDS.
	In \cite{yokota2016smooth}, the tensor completion is based on the low CP rank with the smoothness on the CP components.
	However, the low-rank property and the smoothness are not always consistent through all dimensions. 
	In this work, we seek to estimate the missing entries of a tensor 
	which exhibits the low-rank property in some dimensions, and the smoothness property in the remaining dimensions.

		\vspace{-5mm}
	\section{Motivating Example} \label{sec_motivation_ex}
%
%
%
%
	In this section, we provide a wireless channel learning framework which motivates a noisy tensor completion problem.
	The objective is to provide an approach to portray the map-based wireless channel conditions over the service area.
	In Section \ref{subsec_sys_model}, we introduce the map-based channel model \cite{3gpp.38.901}.
	In Section \ref{subsec_position_aided_beam_recomm}, we introduce the cloud-based position-aided beam recommendation approach.
	In Section \ref{subsec_data_model}, we explain the collection of channel measurements, which are post-processed and recorded along with the side information, constructing a data tensor.
	In Section \ref{subsec_cloud_data_usage}, we describe our proposed cloud-based approach for MIMO beam training. 
	We further investigate the inherent properties of the data measurements, which motivate the design of our proposed tensor completion problem, HNTC. 
			\vspace{-5mm}
	\subsection{Wireless System Model and Problem Description }\label{subsec_sys_model}
\begin{figure}[t]		\centering
		\includegraphics[width=.8\linewidth]{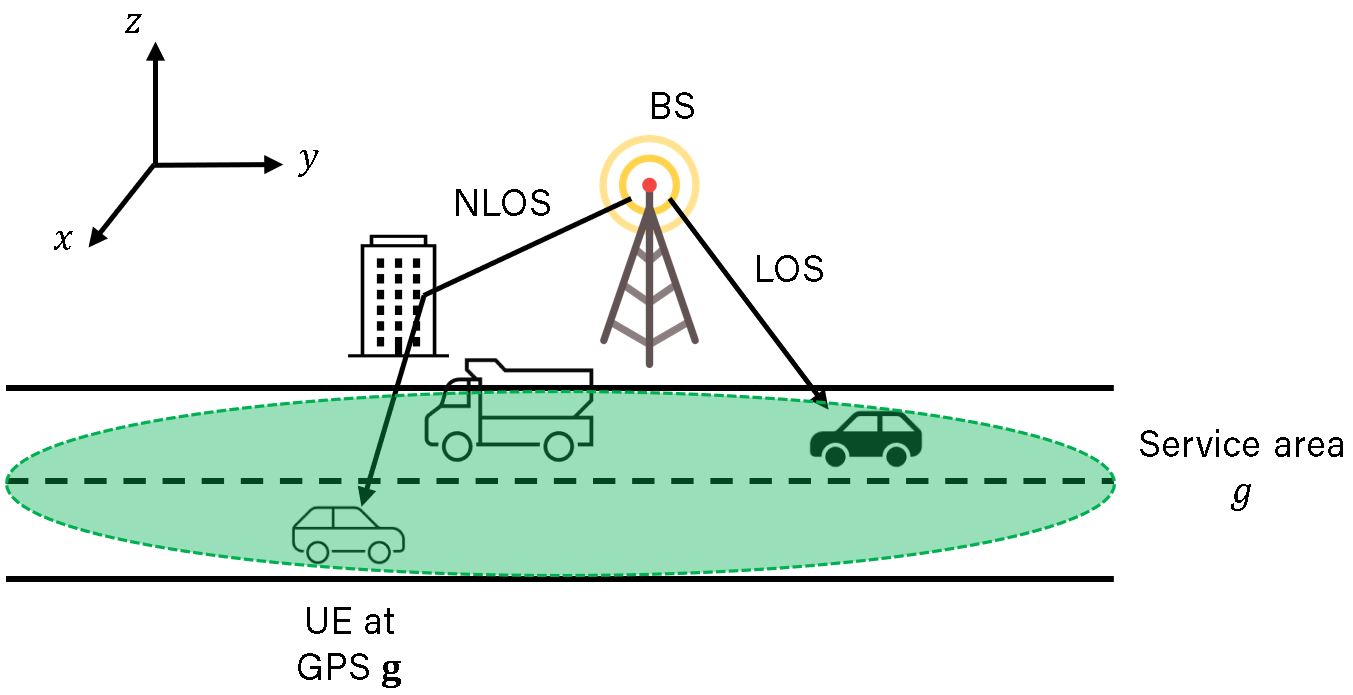}
		\caption{Network layout of the considered scenario. A BS services the mobile UE at GPS $\mathbf{g}$ in the service area $\mathcal G$.}
		\label{fig:wireless_comm_scenario}
		\vspace{-5mm}
	\end{figure}

	We consider a wireless network supporting a geographic region, as depicted in Fig. \ref{fig:wireless_comm_scenario}. 
	A base station (BS) serves mobile UEs in a service area $\mathcal G\subset\mathbb R^2$, modeled as a compact set.
	We assume the BS is at a fixed position and height, and employs an antenna array with $M_r$ antennas with fixed orientation. We consider a reference UE, whose position at time $k$ is denoted as $\mathbf{g}(k)\in\mathcal G$, which employs an antenna array with $M_t$ antennas, with time-varying orientation.
	
	We consider an uplink system to be suitable for data collection since we expect to collect and exploit the measured data from the UE at possible positions in the service area.
	Due to the channel reciprocity, the collected data can also operate in the downlink; in this case, the signal strength received at the UE needs to be fed back to the BS.
	The UE transmits an $M_t\times 1$ signal vector $\mathbf{x}[k]$ at timeslot $k$, yielding the $M_r\times 1$ received signal vector
	\begin{align}
	\label{IO_channel_model}
	\mathbf{y}[k]=\mathbf{H}[k]\mathbf{x}[k] + \mathbf{n}[k],
	\end{align}	
	where $\mathbf{H}[k]$ is an $M_r\times M_t$ narrowband channel matrix of the UE experienced at time $k$, and $\mathbf{n}[k]\sim\mathcal{CN}(\mathbf{0},\sigma_n^2\mathbf{I})$ is an $M_r\times 1$ noise vector. 
	Generally, the channel $\mathbf
	H[k]$ depends on the positions of the BS and UE, the antenna setup (orientation and design) of the BS and UE, and the scattering clusters in the environment and may vary over time as a result of dynamics in the propagation environment, such as fading and mobility of clusters, mobility of the UE and changes in the orientation of the UE's antenna array. 
	However, it exhibits some patterns which depend on the UE position $\mathbf{g}(k)$. 
	In this paper, we are interested in developing a data-aided framework that exploits	these position-dependent patterns to aid the beam-training procedure.

The BS receives the signal with the unit-norm combining vector $\mathbf{w}\in\mathcal W$ taking values from the BS codebook $\mathcal W$,
the UE employs a unit-norm beamforming vector
$\mathbf{f}\in\mathcal F$ taking values from the UE codebook $\mathcal F$.
The transmit signal can then be expressed as $ \sqrt{P_t}\mathbf{f}\cdot\mathbf s$, where $P_t$ is the transmit power, and $\mathbf{s}=[s_1,\dots,s_Q]$ is the known training sequence vector with norm $\sqrt{Q}$.
After signal combining at the receiver with the unit-norm combining vector $\mathbf{w}\in\mathcal W$, the received signal vector can be expressed as
	\begin{align} \label{IO_signal_model_new}
	\Tilde{\mathbf{y}}[k]	=\sqrt{P_t}\mathbf{w}^{H}\mathbf{H}[k]\mathbf{f}\cdot\mathbf{s} + \Tilde{\mathbf n}[k],
	\end{align}	
	where the received noise vector $\Tilde{\mathbf n}[k]\sim \mathcal{CN}(0,\sigma_n^2\mathbf{I})$.
	The received signal power can be expressed as 
	\begin{align}
	\label{eq:rx_pwr}
	r[k] = \lvert  \Tilde{\mathbf y}[k] \mathbf{s}^H \rvert^2 = \lvert \sqrt{P_t}\mathbf{w}^{H}\mathbf{H}[k]\mathbf{f}{\cdot\mathbf s\mathbf s^H} + \hat{n}[k]\rvert^2.
	\end{align}
	where $\hat{n}[k]= \Tilde{\mathbf n}[k] \mathbf s^H$ is a zero-mean complex Gaussian noise with variance $\sigma_n^2{\mathbf s\mathbf s^H}$.

	We assume the use of uniform planar arrays (UPAs) \cite{heath2016overview} at the BS in this work.
	However, the framework presented in this paper can be applied to any codebook design, not just UPAs.
	The UPA codebook can be described by the parameters $(C_y,C_z,C_{\theta},C_{\phi})$, where $(C_y, C_z)$ are the number of antennas in the $y$ and $z$ directions with half wavelength antenna spacing, and $(C_{\theta},C_{\phi})$ are the number of quantized beams along the elevation and azimuth angular directions.
	The array response vector representing a beam pointing in the elevation angle $\theta\in[-\pi/2,\pi/2)$ and the azimuth angle $\phi\in[-\pi/2,\pi/2)$ is denoted as
	\begin{equation}
	\label{atp}
	\begin{aligned}
	\mathbf{a}(\theta,\phi)=\frac{1}{\sqrt{C_yC_z}}
	\begin{bmatrix}
	1\ e^{j\Omega_z}\ \cdots e^{j({C}_z-1)\Omega_z}
	\end{bmatrix}^T
	\otimes\\
	\begin{bmatrix}
	1\ e^{j\Omega_y}\ \cdots e^{j({C}_y-1)\Omega_y}
	\end{bmatrix}^T,
	\end{aligned}
	\end{equation}
	with $\Omega_z{=}{\pi}\sin{\theta}\sin{\phi}$, $\Omega_y{=}{\pi}\sin{\theta}\cos{\phi}$.
	To construct the UPA codebook, $\theta_u$ and $\phi_v$ are uniformly quantized in $[-\pi/2,\pi/2)$ with resolution $\pi/C_{\theta}$ and $\pi/C_\phi$ as
	\begin{align}
	&   \theta_u = -\frac{\pi}{2}+(u-1)\times\frac{\pi}{C_{\theta}}, \ u =1,\dots,C_{\theta},\\
	&  \phi_v = -\frac{\pi}{2}+(v-1)\times\frac{\pi}{C_{\phi}}, \ v =1,\dots,C_{\phi}.
	\end{align}
	We index the beamforming vectors in the codebook as
	$$\mathcal K\equiv\{(u,v):u =1,\cdots,C_{\theta},\ v =1,\cdots,C_{\phi}\},$$
	and we denote the beamforming vector $\mathbf w_{u,v}$ indexed by $(u,v)\in\mathcal K$.

	In this work, we aim to provide a channel learning framework covering the service area in the wireless network.
	With the framework, we are able to provide a set of candidate beams to do the channel estimation for the UE at any possible position.
	To the best of our knowledge, the map-based channel model can only be acquired by real channel measurements, or by simulation via ray-tracing software which simulates the propagation environment.
	To attain our objective upon these facts, we should do the channel measurements or ray-tracing on every possible positions, as assumed in the state-of-the-art \cite{Va2018}, which is impractical in a real system due to the prohibitively large overhead.
	To address the challenge, we propose an efficient data-driven approach to learn the channel conditions of the whole service area with limited number of channel measurements.

	\vspace{-0.5em}
	\subsection{Cloud-Based Position-Aided Beam Alignment}	
	\label{subsec_position_aided_beam_recomm}
	The idea of this approach is to provide a set of candidate beams at a give UE position.
	Since the overhead of the conventional beam-sweeping approach is unacceptable (it scales up with $\lvert\mathcal W\rvert\cdot\lvert\mathcal F\rvert$ and is typically very large),	our objective is to design a learning algorithm that recommends a small
	subset $\mathcal{S}\subset\mathcal W$ of beams to train at the BS, which is likely to contain the best BS beam (the one with the largest received signal power).
	Thus, the training overhead can be reduced to $\lvert\mathcal S\rvert\cdot\lvert\mathcal F\rvert$.
	
\begin{figure}[t]
		\centering
		\includegraphics[width=.8\linewidth]{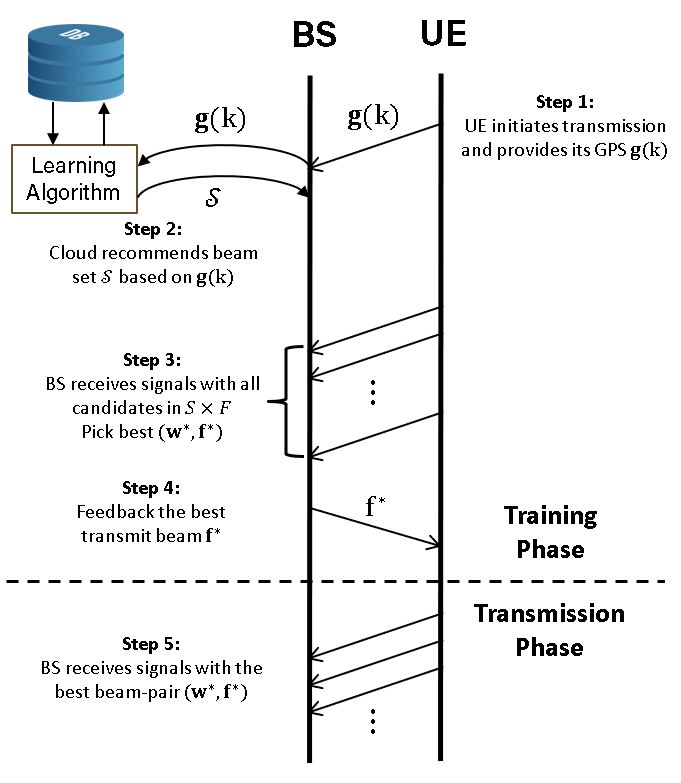}
		\caption{Cloud-based position-aided beam alignment protocol.}
		\label{fig:Beam_recomm_flow}
		 \vspace{-5mm}
	\end{figure}

	In Fig. \ref{fig:Beam_recomm_flow}, we introduce the flow diagram of the cloud-based position-aided beam alignment.
	In Step 1, the UE initiates the uplink transmission request with its current GPS coordinate $\mathbf{g}(k)$ to the BS using sub-6 GHz control channels.
	This information is available via a suite of sensors such as GPS \cite{roy2014smartphone,maschietti2017robust}.
	In Step 2, the BS forwards the GPS coordinate $\mathbf{g}(k)$ to the cloud, which processes the learning	algorithm to provide the recommended beam set $\mathcal{S}$ to train at the BS.
	Note that, in order to provide robustness against
	different UE antenna designs (such as UPAs with different number of antennas and configurations)
	and temporal dynamics in UE antenna orientation, our proposed  algorithm
	recommends the beam-training set $\mathcal S$ only on the BS side (whose antenna has a fixed position and orientation), but
	 does not recommend a beam-training set on the UE side, so that the UE is required to scan all possible beamforming vectors in $\mathcal{F}$ to acquire the best one to be used for data communication. 
	Feedback-based beam-training schemes such as \cite{6600706,hussain2018energy,9013578}, or schemes that leverage the mobility of the UE such as \cite{hussain2020mobility}, can be used to further reduce the training overhead, but the analysis of this case is out of the scope of this paper.	
	In Step 3, the UE transmits a sequence of $\lvert\mathcal{S}\rvert\cdot \lvert\mathcal{F}\rvert$ known signals, where the BS receives the signals with the beam pairs  \textbf{$\{(\mathbf{w},\mathbf{f}):\mathbf{w}\in\mathcal{S},\ \mathbf{f}\in\mathcal{F}\}$}.
	Then, the BS selects the best $(\mathbf{w}^*,\mathbf{f}^*)$ among all candidate beam pairs ranked by the received power.
	In Step 4, the BS feeds back the selected UE beamforming vector $\mathbf{f}^*$ to the UE.
	In Step 5, the UE communicates with the BS using the selected beam pair $(\mathbf{w}^*,\mathbf{f}^*)$ for the subsequent uplink or downlink data transmission.

	\vspace{-5mm}
	\subsection{Data Model}\label{subsec_data_model}
	Here, we describe how we collect the channel measurements, and store the information in the database.
	Let $\bar{\mathcal G}=[X_0,X_{end}]\times[Y_0,Y_{end}]$ be a rectangular region containing the service area $\mathcal G$.
	We discretize $\bar{\mathcal G}$ with resolution $\Delta_s$, thus defining the discrete GPS coordinates $\mathbf g = (g_x,g_y)\in\bar{\mathcal G}$.
	Then, we define the position labels $\mathbf p = (p_x,p_y)$ with $p_x\in \{1,\cdots,L_x\}$ and $p_y\in \{1,\cdots,L_y\}$, where $L_x = \left\lceil\frac{X_{end}-X_0}{\Delta_s}\right\rceil$ and $L_y =\left\lceil \frac{Y_{end}-Y_0}{\Delta_s}\right\rceil$ are the number of points in the $x$ and $y$ axes, and $\left\lceil x \right\rceil$ is the ceiling function.
	The function $\rho(\mathbf{g})$ maps the coordinate $\mathbf g\in \bar{\mathcal G}$ to the closest discretized position label $\mathbf p$ as
	\begin{equation}
	\label{pos_label_derive}
	\mathbf p =\rho(\mathbf{g})=\left(1 + \round{\frac{g_x-X_0}{\Delta_s}},1 + \round{\frac{g_y-Y_0}{\Delta_s}}\right),
	\end{equation}
	where $\round{x}$ denotes the nearest integer to $x$.	
	
	During the data collection, the BS measures the received power, using the UE's current position and beam as side information.
	To do so, the BS receives the signal using an arbitrary combining vector  $\mathbf{w}_{u,v}\in\mathcal{W}$ indexed by $(u,v)\in\mathcal K$, whereas the UE  performs the beam-training exhaustively over the UE beamforming set $\mathcal F$;
	with the UE in the discretized position $\mathbf p=\rho(\mathbf{g}(k))$,
	the strongest received signal power is then denoted as 
	\begin{equation} \label{eq_rx_pwr_data}
	r_{k}^{(\mathbf{p},u,v)}=\max_{\substack{ \mathbf{f}_i\in\mathcal{F}\\i=0,\dots,\vert\mathcal F\vert-1}}\lvert\sqrt{P_t}\mathbf{w}_{u,v}^H\mathbf{H}[k+i]\mathbf{f}_i+\hat{n}[k+i]\rvert^2,
	\end{equation}
	where $\mathbf{H}[k]$ is the uplink MIMO channel of the UE at timeslot $k$, and $\hat{n}[k]$ is the received noise at timeslot $k$.
	We assume that, during this process, the discretized UE position does not change, i.e. $\rho(\mathbf g(k+i))$ remains fixed $\forall i=0,\dots,\vert\mathcal{F}\vert -1$.
	The received signal power $r_k^{(\mathbf{p},u,v)}$ at timeslot $k$, along with the side information $\mathbf p$ (UE discretized position) and $(u,v)$ (BS beamforming index), is then recorded in the database, with the procedure described below.

	The BS might collect multiple measurements on a given combination of position $\mathbf p$ and combining vector $\mathbf{w}_{u,v}$.
	Therefore, the cloud database stores an average of the measurements collected. Moreover, to enable adaptation in non-stationary environments, it computes a weighted average which discounts past measurements. To this end, let $\chi_k^{(\mathbf{p},u,v)}\in\{0,1\}$ be an indicator variable, equal to one if and only if 
	a  measurement is collected at time $k$ in the discretized position $\mathbf p$ on the combining vector $\mathbf{w}_{u,v}$.
	Let $\bar r_k^{(\mathbf{p},u,v)}$ be the corresponding weighted received power stored in the database at time $k$ (initially, $\bar r_0^{(\mathbf{p},u,v)}=0$). This is computed as
	\begin{align}
	\label{avg_rx_power}
	\bar r_k^{(\mathbf{p},u,v)}=\frac{\sum_{\ell=0}^k\alpha^\ell\chi_{k-\ell}^{(\mathbf{p},u,v)}r_{k-\ell}^{(\mathbf{p},u,v)}}{\sum_{\ell=0}^k\alpha^\ell\chi_{k-\ell}^{(\mathbf{p},u,v)}},
	\end{align}
	where $\alpha\in(0,1]$ is a discount factor which enables adaptation to non-stationary environments. In the stationary-case, one can set $\alpha=1$ and \eqref{avg_rx_power} yields the sample average of past measurements.
	In addition to $\bar r_k^{(\mathbf{p},u,v)}$, the database stores also the weighted number of past measurements
	collected on a given position and beam, denoted as $\bar N_k^{(\mathbf{p},u,v)}$ and given by
	\begin{align}
		\label{avg_N}
		\bar N_k^{(\mathbf{p},u,v)}=\sum_{\ell=0}^k\alpha^\ell\chi_{k-\ell}^{(\mathbf{p},u,v)},
	\end{align}
	which yields the total number of measurements collected up to time $k$ in the stationary case ($\alpha=1$).
	In fact, the average received power $\bar r_k^{(\mathbf{p},u,v)}$ is a noisy estimate of the expected signal power, and the variance of this measurement typically decreases with the number of measurements. In case of $N$ i.i.d. measurements, the variance decreases by a factor $1/N$. 
	Therefore, the term $\bar N_k^{(\mathbf{p},u,v)}$ will be used to track the accuracy of the received power estimates stored in the database, and will be used in the noisy tensor completion algorithm developed in the next section.

	Note that \eqref{avg_rx_power} and \eqref{avg_N} can be computed in an online fashion, without requiring to store all past measurements. In fact, upon collecting a new measurement $r_{k+1}^{(\mathbf{p},u,v)}$ in position $\mathbf{p}$ using the combiner $\mathbf{w}_{u,v}$ at timeslot $k+1$, \eqref{avg_rx_power} and \eqref{avg_N} can be updated as
	\begin{align}
	&\bar N_{k+1}^{(\mathbf{p},u,v)}=\alpha\bar N_{k}^{(\mathbf{p},u,v)}+\chi_{k+1}^{(\mathbf{p},u,v)},
	\\
	&\bar r_{k+1}^{(\mathbf{p},u,v)}=\bar r_{k}^{(\mathbf{p},u,v)}+\frac{\chi_{k+1}^{(\mathbf{p},u,v)}}{\bar N_{k+1}^{(\mathbf{p},u,v)}}\left(r_{k+1}^{(\mathbf{p},u,v)}-\bar r_{k}^{(\mathbf{p},u,v)}\right),
	\end{align}
	which are updated in the database along with the side information, including the UE position $\mathbf{p}$ and the index of BS combining vector $(u,v)$, as in TABLE \ref{table:database}.
	\vspace{-5mm}
	\begin{table}[h]
		\caption{Database form} 
		\label{table:database} 
		\centering 
		\begin{tabular}{|c c c c   | c|c|} 
			\hline 
			$p_x$ & $p_y$ & $u$ & $v$ & $\Bar{r}^{(\mathbf{p},u,v)}$ & $\bar{N}^{(\mathbf{p},u,v)}$ \\ [0.5ex] 
			\hline 
			1 & 1  & 1 & 4 & 5.2 & 3\\
			1 & 2  & 4 & 5 & 6.1 & 1\\
			$\vdots$ & $\vdots$ & $\vdots$& $\vdots$& $\vdots$& $\vdots$\\ [1ex] 
			\hline 
		\end{tabular}
		\vspace{-3mm}
	\end{table}

	We represent the average receive power as a fourth order tensor in $\mathbb{R}^{L_x\times L_y\times C^r_{\theta}\times C^r_{\phi}}$:
	\begin{align}\label{eq_data_tensor}
	&\mathcal{T}(\mathbf{p},u,v)
	=\left\{
	\begin{array}{ll}
	\Bar{r}^{(\mathbf{p},u,v)},&(\mathbf{p},u,v)\in\Psi,\\
	0, &\mathrm{otherwise,}\\
	\end{array}
	\right.
	\end{align}
	where $\Psi$ is the set of observed combinations of positions and BS codewords stored in the database and the unobserved entries $(\mathbf{p},u,v)\notin\Psi$ are set to zero.
	Moreover, the weighted number of measurements for all positions and BS codewords are recorded as a fourth order tensor in $\mathbb{R}^{L_x\times L_y\times C^r_{\theta}\times C^r_{\phi}}$:
	\begin{equation} \label{eq_data_tensor_measurements}
	\mathcal{V}(\mathbf{p},u,v) = \bar N^{(\mathbf{p},u,v)},
	\end{equation}
	which can be updated in an online fashion as new measurements are collected, as described previously.
    It is impractical to collect the measurements with all combinations of positions and BS codewords into the database due to the limited sampling resource and non-stationarity of the propagation environment which causes past measurements to become outdated.
    Therefore, some positions possibly have no representation in the database.
    Even in the observed positions, there might be only a limited number of BS codewords' information recorded.
    Thus, the data tensor $\mathcal{T}$ may be highly incomplete.


	\vspace{-0.8em}

	\subsection{What do we do with the data?} \label{subsec_cloud_data_usage}
	Our goal is to recommend a set of $N_{tr}$ candidate beams for the BS to train based on UE position.
	If the UE is in a position $\mathbf{p}$ represented in the database, and the measurements with all beams available, then this task can be easily accomplished by recommending the $N_{tr}$ beams with highest average received power in the given position.
	Otherwise, we are required to design an algorithm based on tensor completion that employs the knowledge of limited number of beams' information at UE's neighboring positions to provide the beam recommendation.

	To address the issue, given the data tensor $\mathcal{T}$ \eqref{eq_data_tensor} and the measurement tensor $\mathcal{V}$ \eqref{eq_data_tensor_measurements}, we formulate the noisy tensor completion problem as
	\begin{align}
	\label{hybrid_tensor_completion_problem_ori}
	&\!\!\arg\min_{\mathcal{X}}\  \sum_{p_x,p_y}\lVert 
	\mathcal{X}(p_x,p_y,:,:)\rVert_*
	+\gamma \sum_{u,v}{{S}}\left(\mathcal{X}(:,:,u,v)\right)\!\!\\	
	\label{sqerr_ori}
	&\mathrm{s.t.}\ \sum_{\psi} \mathcal{W}(\psi)\left(\mathcal{T}(\psi) - \mathcal{X}(\psi)\right)^2 \leq \eta,	
	\end{align}
	where $\gamma$ is the trade-off parameter between the smoothness property across the positional dimensions and the low-rank property across the beam dimensions.
	The smoothness function $S(\mathcal{X})$ measures the difference between the values of adjacent elements in $\mathcal{X}$.
	The minimizer tensor $\mathcal{X}^*$ is the reconstructed tensor.
	The tensor $\mathcal{W}$ is the weighting tensor, with elements $\mathcal{W}(\psi)={\mathcal{V}(\psi)}/{\sum_{\psi^'}\mathcal V(\psi^')}$ associated to the entry $\psi$.
	For an unobserved element $\psi$, $\mathcal{W}(\psi)=0$, so that the corresponding entry has no contribution to the error term \eqref{sqerr_ori}.	
	The inequality constraint {\eqref{sqerr_ori}} originates from the noisy measurements, and induces an upper bound $\eta$ on the weighted sum of the mean square error between $\mathcal{X}$ and $\mathcal{T}$, which provides additional flexibility to the observed elements. 
	The noiseless tensor completion is a special case when $\eta=0$. Note that
	tensor entries associated to more measurements (larger $\mathcal V(\psi)$) are associated a larger weight ($\mathcal W(\psi)$), so they contribute more to the squared error term \eqref{sqerr_ori}. In fact, this is a desirable feature since the variance of the sample averaged receive power decreases with the number of measurements.
	
	The first term of the objective function \eqref{hybrid_tensor_completion_problem_ori} represents the low-rank property of the measurement data.
	For each position, the channel sparsity of mmWave propagation dictates that only few combinations of beams lead to significant signal measurements, which imposes the low-rank property to the collected measurement data tensor in the beam dimensions.
	However, the tensor completion with low-rank alone often fails to provide predictions on unobserved positions due to the dependence on the sampling set, especially in the case with high ratio of missing elements.
	The smoothness can be considered to support the data recovery.
	Regarding the second term of the objective function \eqref{hybrid_tensor_completion_problem_ori},
	it quantifies the smoothness of the measurement data.
	The spatial coherence of the channels in adjacent positions can be exploited to learn the channel conditions.
	The received power measured using a certain beam exhibits spatial correlation, so that the corresponding data tensor exhibits smoothness across the positional dimensions.
	
	
	The low-rank property across the beam dimensions and the smoothness property across the positional dimensions will be exploited to develop the HNTC algorithm in the next section.

	\vspace{-0.5em}
	\section{Hybrid Noisy Tensor Completion}\label{sec_proposed_TC}
	In the previous section, we introduced a data-driven approach to capture the channel conditions by constructing a data tensor as in \eqref{eq_data_tensor}, containing the position and BS codeword as the side information.
	In a wireless network, the UEs are possibly equipped with different antenna designs.
	To enable our proposed approach on a variety of UE antenna designs, we construct the data tensor by collecting the received signal power under the best UE beam.
	Besides, the channel is time-varying in the propagation environment, so we record the weighted average received power in the data tensor to learn some structures preserved over time.
	However, due to the limited sampling resources, the collected data tensor might be highly incomplete, so it is crucial to devise an algorithm for the tensor completion problem	that exploits both low-rank and smoothness properties across separate dimensions.
	The noise effect is also an important factor to be considered since the signal in wireless systems is inevitably contaminated. 
	
	Section \ref{TC_problem} formulates a noisy tensor completion problem, and then proposes the algorithm, HNTC, to solve the tensor completion problem using the ADMM method. 
	Section \ref{Online_TC_algorithm} proposes the online HNTC, and Section \ref{complexity} discusses the computational complexity.
	
	\subsection{Proposed Noisy Tensor Completion} \label{TC_problem}
	We propose a completion problem for recovering an incomplete data tensor, using the assumption that a portion of the dimensions have the smoothness property and the remaining dimensions are low-rank.
	As motivated in the previous section, we consider a tensor $\mathcal{T}{\in}\mathbb{R}^{I^s_{1}{\times}\cdots\times I^s_{n_1}\times I^{\ell}_{1}\times\cdots\times I^{\ell}_{n_2}}$ which separates its dimensions into two partitions. 
	The first $n_1$ dimensions are smooth and the last $n_2$ dimensions have the low-rank property.
	It means that the $n_1$-th order subtensor obtained by fixing the indices of the last $n_2$ dimensions, $\mathcal{T}(:,\cdots,:,\mathbf i_{1: n_2}) \in\mathbb{R}^{I^s_{1}\times\cdots\times I^s_{n_1}}$, is a smooth tensor;
	and the $n_2$-th order subtensor obtained by fixing the indices of the first $n_1$ 
	dimensions, 
	$\mathcal{T}(\mathbf i_{1:n_1} ,:,\cdots,:) \in\mathbb{R}^{I^{\ell}_{1}\times\cdots\times I^{\ell}_{n_2}}$, is low-rank.	
	We use the following smoothness metric.
	\begin{definition}
		\label{def_smooth_func}
		The linear tensor total variation (LTTV) of the tensor 
		$\mathcal{B}\in\mathbb{R}^{I_1\times \cdots \times I_M}$, quantifying 
		the smoothness of the tensor $\mathcal{B}$, is defined as 
		$$
		\mathrm{LTTV}(\mathcal{B})
		=\sum_{\mathbf i_{1: m}}\sum_{d=1}^{M}\left[\delta_d(\mathcal{B}(\mathbf i_{1: m}))\right]^2,$$
		where
		$\delta_d(\mathcal{B}(\mathbf i_{1: M}))=\mathcal{B}(\mathbf i_{1: M}+\mathbf e_d)-\mathcal{B}(\mathbf i_{1: M})$, $\mathbf e_d$ is the vector of zeros except in the $d$-th position in which it is equal to 1.
	\end{definition}
	Note that $\mathrm{LTTV}(\mathcal{B})$  is a natural extension of
	linear total variation (LTV) \cite{han2014linear} to tensors, which accumulates the element squared differences of $\mathcal{B}$ along every dimension.

	Assuming an incomplete tensor $\mathcal{T}$ and the associated tensor $\mathcal{V}$ representing the number of measurements collected on each entry,
	the goal is	to estimate the missing elements in $\mathcal{T}$ by solving
	\begin{align}
	\label{hybrid_tensor_completion_problem}
	&\arg\min_{\mathcal{X}}\  {	\sum_{\mathbf i_{1: n_1}}\sum_{k=1}^{n_2}\alpha_k\left\Vert 
	\left\{\mathcal{X}{(\mathbf i_{1:n_1},:,\cdots,:)}\right\}_{(k)}\right\Vert_*}\\
	\nonumber
	&\hspace{2em}+\gamma 
	\sum_{\mathbf i_{1:n_2}}{\mathrm{LTTV}}\left(\mathcal{X}(:,\cdots,:,\mathbf i_{1: n_2})\right)\\
	&\mathrm{s.t.}\ \sum_{\psi} \mathcal{W}(\psi)\left(\mathcal{T}(\psi) - \mathcal{X}(\psi)\right)^2 \leq \eta,
	\label{sqerr}
	\end{align}
	where $\gamma$ is the trade-off parameter between the smoothness property across the first $n_1$ dimensions and the low-rank property across the remaining $n_2$ dimensions, and the constraint \eqref{sqerr} accounts for the noise in the tensor $\mathcal T$.
	Note that this formulation employs the tensor nuclear norm defined in \eqref{def_tensor_nuclear_norm}, and the term $\left\{\mathcal{X}{(\mathbf i_{1:n_1},:,\cdots,:)}\right\}_{(k)}$ is the mode-k unfolding of $\mathcal{X}{(\mathbf i_{1:n_1},:,\cdots,:)}\in\mathbb{R}^{I^{\ell}_{1}\times\cdots\times I^{\ell}_{n_2}}$.
	The tensor $\mathcal{W}$ is the weighting tensor, with elements $\mathcal{W}(\psi)={\mathcal{V}(\psi)}/{\sum_{\psi^'}\mathcal V(\psi^')}$ associated to the entry $\psi$.
	The minimizer tensor $\mathcal{X}^*$ is the reconstructed tensor.
	The proposed HNTC optimization problem is convex, since it is a linear combination of 
	the tensor nuclear norm (convex, see \eqref{opt_prob_Liu})
	and the LTTV (quadratic and convex, see \textbf{Definition} \ref{def_smooth_func}), and the inequality constraint is also a convex quadratic function.
	The problem \eqref{hybrid_tensor_completion_problem} is a general form of the tensor completion problem in \eqref{hybrid_tensor_completion_problem_ori}, where the positional information are associated with the smooth dimensions, and the beam information are associated with the low-rank dimensions. 
	Note that the proposed HNTC problem is a generalization of other state-of-the-art tensor completion problems. 
	By setting $n_1=0$, we obtain the low-rank tensor completion \cite{6138863};	by setting $n_2=0$, we obtain the  linear tensor total variation minimization of \cite{han2014linear}.

		\begin{figure*} [hb!]
		\hrulefill	
		\begin{equation} \tag{35}
		\label{Update_A}
		\begin{aligned}
		&\mathcal{A}^{\mathbf i_{1:n_1}}(\mathbf{\hat i}_{1:n_1})=
		&\left\{
		\begin{array}{ll}
		{n_2 \cdot\lambda}+2\gamma\sum_{k=1}^{n_1}\{\delta(i_k>1)+\delta(i_k<I^s_k)\}
		&,\ \mathrm{ if }\ \mathbf{\hat i}_{1:n_1}=\mathbf i_{1:n_1},\\
		-2\gamma &,\ \mathrm{ if }\ \lVert \mathbf{\hat i}_{1:n_1}-\mathbf i_{1:n_1}\rVert_2=1,\\
		0 &,\ \mathrm{ otherwise}.
		\end{array}
		\right.\\
		\end{aligned}
		\end{equation}
	\end{figure*}
	
	We use the ADMM technique \cite{boyd2011distributed} to solve \eqref{hybrid_tensor_completion_problem}.
	The optimization problem is reformulated as
	\begin{align}
	\label{Lagrange_form_optimization_problem}
	&\arg\ \min_{\mathcal{X},\{\mathcal{Y}_k\}_{k=1}^{n_2}}\  
	\sum_{\mathbf i_{1: n_1}}\sum_{k=1}^{n_2}\alpha_k\left\Vert 
	\left\{\mathcal{Y}_k{(\mathbf i_{1:n_1},:,\cdots,:)}\right\}_{(k)}\right\Vert_*\\
	\nonumber
	&\hspace{0em}+\gamma 
	\sum_{\mathbf i_{1:n_2}}\mathrm{LTTV}\left(\mathcal{X}(:,\cdots,:,\mathbf i_{1:n_2})\right) 
	\hspace{0em}+\sum_{k=1}^{n_2}{\frac{\lambda}{2}\left\Vert(\mathcal{Y}_k-\mathcal{X})\right\Vert_F^2}\\
	&\mathrm{s.t.}\ 
	\label{ineqconst}
	\sum_{\psi}{\mathcal W(\psi)}(\mathcal{T}(\psi) - \mathcal{X}(\psi))^2 \leq \eta,\\
	&\hspace{1.5em}	\mathcal{X} = \mathcal{Y}_k,\ k=1,\dots,n_2,
	\label{eqconst}
	\end{align}
	where $\lambda$ is a small fixed positive parameter.
	With ADMM, we introduce the variables $\{\mathcal{Y}_k\}_{k=1}^{n_2}$ to separate the smooth and low-rank dimensions of the tensor.
	The additional equality constraints $\mathcal{X}=\mathcal{Y}_k,\ k=1,\dots,n_2$ guarantee that \eqref{Lagrange_form_optimization_problem} is equivalent to the problem \eqref{hybrid_tensor_completion_problem}.
	
	We introduce the Lagrangian multiplier $\mathcal{Z}_k$ 
	associated with the $k$-th equality constraint \eqref{eqconst},
	and $\mu$ associated with the inequality constraint \eqref{ineqconst}.
	The corresponding augmented Lagrangian function is expressed as
	\begin{align}
	\label{Aug_Lagrange_func}
	\nonumber
	L&\Big({\mathcal{X},\{\mathcal{Y}_k\}_{k=1}^{n_2},\{\mathcal{Z}_k\}_{k=1}^{n_2},\mu}\Big)\\	 
	\nonumber
	=& \sum_{k=1}^{n_2}\Bigg\{ 
	\sum_{\mathbf i_{1: n_1}}\alpha_k\left\Vert 
	\left\{\mathcal{Y}_k{(\mathbf i_{1:n_1},:,\cdots,:)}\right\}_{(k)}\right\Vert_*\\
	\nonumber
	&\hspace{2.5em}
	+\bigg<\mathcal{Z}_k,\mathcal{Y}_k-\mathcal{X}\bigg>
	+\frac{\lambda}{2}\left\Vert\mathcal{Y}_k-\mathcal{X}\right\Vert_F^2\Bigg\}
	\\
	\nonumber
	&+\gamma 
	\sum_{\mathbf i_{1:n_2}}\mathrm{LTTV}\left(\mathcal{X}(:,\cdots,:,\mathbf i_{1:n_2})\right)\\
	&+\mu \left(\sum_{\psi}\mathcal{W}(\psi)(\mathcal{T}(\psi) - \mathcal{X}(\psi))^2 - \eta\right).
	\end{align}
	The ADMM algorithm is implemented by minimizing $L$	over $\mathcal{X},\mathcal{Y}_k$ and then updating the 
	Lagrangian multipliers $\mathcal{Z}_k$ and $\mu$ in an iterative fashion as 
	\begin{align}
	\label{update_overview}
	&\mathcal{X}_{t+1}=\arg\min_{\mathcal X} 
	L\left({\mathcal{X},\{\mathcal{Y}_{k,t}\}_{k=1}^{n_2},\{\mathcal{Z}_{k,t}\}_{k=1}^{n_2},\mu_t}\right);\\
	\nonumber
	&\mathcal{Y}_{k,t+1}= 
	\arg\min_{\mathcal{Y}_k} 
	L\left({\mathcal{X}_{t+1},\{\mathcal{Y}_k\}_{k=1}^{n_2},\{\mathcal{Z}_{k,t}}\}_{k=1}^{n_2},\mu_t\right),\\ 
	&\hspace{4.1em}\forall k=1,\dots,n_2;\\
	\label{update_Z}
	&\mathcal{Z}_{k,t+1}=\mathcal{Z}_{k,t}+{\beta_1}\left(\mathcal{Y}_{k,t+1}-\mathcal{X}_{t+1}\right),\ \forall k=1,\dots,n_2;\\ 
	\label{update_nu}
	&\mu_{t+1}=\left(\mu_t + \beta_2 \Big(\sum_{\psi}\mathcal{W}(\psi)(\mathcal{T}(\psi) - \mathcal{X}_{t+1}(\psi))^2- \eta\Big)\right)^+;
	\end{align}
	where $\beta_1$ is the step-size for updating $\mathcal{Z}_k$, $\beta_2$ is the step-size for updating $\mu$, and $(\mu)^+=\max(\mu,0)$ is the projection of $\mu$ onto $\mathbb{R}_+$.


	To optimize $\mathcal{X}$, we minimize $L$ 
	with fixed $\{\mathcal{Y}_{k,t}\}_{k=1}^{n_2}$, $\{\mathcal{Z}_{k,t}\}_{k=1}^{n_2}$, and $\mu_t$, yielding
	\begin{align}\label{X_subproblem}
		\nonumber
		{\mathcal X_{t+1}=}		&\arg\min_{\mathcal{X}}\ \sum_{k=1}^{n_2}\left\{\frac{\lambda}{2}\lVert\mathcal{Y}_{k,t}-\mathcal{X}\rVert_F^2+\big<\mathcal{Z}_{k,t},\mathcal{Y}_{k,t}-\mathcal{X}\big>\right\}\\
		\nonumber
		&\hspace{3.5em}+\gamma 
		\sum_{\mathbf i_{1:n_2}}\mathrm{LTTV}\left(\mathcal{X}(:,\cdots,:,\mathbf i_{1:{n_2}})\right)\\
		&\hspace{3.5em}+\mu_t \sum_{\psi}\mathcal{W}(\psi)(\mathcal{T}(\psi) - \mathcal{X}(\psi))^2 .
	\end{align}
	To solve \eqref{X_subproblem}, we consider the optimization problem for each $\mathbf i_{1:{n_2}}$, separately.
	For a given $\mathbf i_{1:{n_2}}$, we define the subtensor $\hat{\mathcal{X}}=\mathcal{X}(:,\cdots,:,\mathbf i_{1:{n_2}})$, 
	$\hat{\mathcal{Y}}_{k,t}=\mathcal{Y}_{k,t}(:,\cdots,:,\mathbf i_{1:{n_2}})$,
	$\hat{\mathcal{Z}}_{k,t}=\mathcal{Z}_{k,t}(:,\cdots,:,\mathbf i_{1:{n_2}})$.
	For the data tensor, we define the subtensor $\hat{\mathcal{T}}=\mathcal{T}(:,\cdots,:,\mathbf i_{1:{n_2}})$, $\hat{\mathcal{W}}=\mathcal{W}(:,\cdots,:, \mathbf i_{1:{n_2}})$.
	Then, the optimization problem can be reformulated as
	\begin{align} \label{sub_X_prob_fix_beam}		
	&\arg\min_{\hat{\mathcal{X}}}\ \sum_{k=1}^{n_2}\left\{\frac{\lambda}{2}\lVert\hat{\mathcal{Y}}_{k,t}-\hat{\mathcal{X}}\rVert_F^2+\big<\hat{\mathcal{Z}}_{k,t},\hat{\mathcal{Y}}_{k,t}-\hat{\mathcal{X}}\big>\right\}\\
	\nonumber
	&\hspace{3.5em}+\gamma \mathrm{LTTV}\left(\hat{\mathcal{X}}\right)+\mu_t
	\sum_{\psi}\mathcal{\hat W}(\psi)(\mathcal{\hat T}(\psi) - \mathcal{\hat X}(\psi))^2 .
	\end{align}
	Since the objective function is a quadratic function of $\hat{\mathcal{X}}$, we compute the derivative of \eqref{sub_X_prob_fix_beam} with respect to each element in $\hat{\mathcal{X}}$ and set it equal to zero, yielding the system of equations
	\begin{align}\label{Noisy_X_unknown_eq}
	\nonumber
	&\left<\mathcal{A}^{\mathbf i_{1:n_1}},\hat{\mathcal{X}}\right>+2{\mu_t}\mathcal{\hat W}(\mathbf i_{1:n_1})\left(\mathcal{\hat X}(\mathbf i_{1:n_1})-\mathcal{ \hat T}(\mathbf i_{1:n_1})\right)\\
	&={\sum_{k=1}^{n_2}\left\{\lambda\mathcal{\hat Y}_{k,t}(\mathbf i_{1:n_1})+\mathcal{\hat Z}_{k,t}(\mathbf i_{1:n_1})\right\}},\ \forall \mathbf i_{1:n_1},
	\end{align}
	where $\mathcal{A}^{\mathbf i_{1:n_1}}\in\mathbb{R}^{I^s_1\times \cdots\times I^s_{n_1}}$ is defined in \eqref{Update_A}.
	There are $\prod_{d=1}^{n_1}I^s_d$ unknowns and $\prod_{d=1}^{n_1}I^s_d$ linear equation, so that the subtensor $\hat{\mathcal{X}}$ can be found by solving the linear system in \eqref{Noisy_X_unknown_eq}.
	Note that the linear independence of the system of equations depends on the measurement data and may not be guaranteed, so that the solution may not be unique (in this case, it can be found using the Moore-Penrose pseudo-inverse).

	The minimization of $L$ over $\mathcal{Y}_k$ with fixed $\mathcal{X}_{t+1}$, $\{\mathcal{Z}_{k,t}\}_{k=1}^{n_2}$, and $\mu_t$ can be formulated as
	\begin{equation}
	\begin{aligned}
	\nonumber
	&\mathcal{Y}_{k,t+1}=\arg\min_{\mathcal{Y}_k}\ \sum_{\mathbf i_{1: n_1}}
	\Bigg\{{\alpha_k}\left\Vert 
	\left\{\mathcal{Y}_k{(\mathbf i_{1: n_1},:,\cdots,:)}\right\}_{(k)}\right\Vert_*\\
	&+\frac{\lambda}{2}\Big\Vert\bigg\{\Big\{\mathcal{Y}_k    
	-\Big(\mathcal{X}_{t+1}-\frac{\mathcal{Z}_{k,t}}{\lambda}\Big)\Big\}(\mathbf i_{1:n_1},:,\cdots,:)\bigg\}_{(k)}\bigg\Vert_F^2\Bigg\}.\\
	\end{aligned}
	\end{equation}
	For a given $\mathbf i_{1: n_1}$, we define 
	$\mathcal{\tilde{X}}_{t+1}=\mathcal{X}_{t+1}{(\mathbf i_{1: n_1},:,\cdots,:)}$,	$\mathcal{\tilde{Y}}_k=\mathcal{Y}_k{(\mathbf i_{1: n_1},:,\cdots,:)}$, and $\mathcal{\tilde{Z}}_{k,t}=\mathcal{Z}_{k,t}{(\mathbf i_{1: n_1},:,\cdots,:)}$.
	Thus, we reformulate the problem as
	\begin{equation*}
		\arg\min_{\mathcal{\tilde Y}_k}\ 
		{\alpha_k}\left\Vert 
		\left\{\mathcal{\tilde Y}_k\right\}_{(k)}\right\Vert_*
		+\frac{\lambda}{2}\bigg\Vert\bigg\{\mathcal{\tilde Y}_k    
		-\Big(\mathcal{\tilde X}_{t+1}-\frac{\mathcal{\tilde Z}_{k,t}}{\lambda}\Big)\bigg\}_{(k)}\bigg\Vert_F^2.
	\end{equation*}
	This problem is shown to be strictly convex \cite{cai2010singular}, and the solution is given by singular value thresholding.
	The update can be written as
	\setcounter{equation}{35}
	\begin{equation} \label{Y_sol} 
		\mathcal{\tilde Y}_k= \text{fold}_k\left(\mathcal{D}_{\frac{\alpha_k}{\lambda}}\left(\left\{\mathcal{\tilde X}_{t+1}-\frac{\mathcal{\tilde Z}_{k,t}}{\lambda}\right\}_{(k)}\right)\right),
	\end{equation}
	where $\mathcal{D}_\tau$ is the soft-thresholding operator.
	For a matrix $\mathbf A$ with singular value decomposition (SVD) 
	$\mathbf{A}=\mathbf{U}\mathbf{\Sigma}\mathbf{V}^H$,
	where $\mathbf{\Sigma}=\text{diag}(\sigma_1,\dots\sigma_r)$, this operation is defined as
	$\mathcal{D}_\tau(\mathbf{A})=\mathbf{U}\mathcal{D}_\tau(\mathbf{\Sigma})\mathbf{V}^H,\	 
	\mathcal{D}_\tau(\mathbf{\Sigma})=\text{diag}(\{\max\{\sigma_i-\tau,0\}\}).$
	With the update of $\tilde{\mathcal{Y}}_k=\mathcal{Y}_k(\mathbf i_{1: n_1},:,\cdots,:)$ for each $\mathbf i_{1: n_1}$, the updated $\mathcal{Y}_{k,t+1}$ can thus be completed.
	
	
	Then, we update the Lagrangian multipliers $\mathcal{Z}_k$ and $\mu$ as in \eqref{update_Z} and \eqref{update_nu}.
	With the convergence of ADMM \cite{boyd2011distributed}, the iteration approaches the primal feasibilities in \eqref{ineqconst} and \eqref{eqconst},	and we set the stop criteria as 
	\begin{equation}\label{stop_criterion_1}
		\sum_{k=1}^{n_2}\left\Vert\mathcal{X}_t-\mathcal{Y}_{k,t}\right\Vert_F< \epsilon,
	\end{equation}	
	where $\epsilon$ is a constant threshold, and 
	\begin{equation}\label{stop_criterion_2}
	\sum_{\psi}\mathcal{W}(\psi)(\mathcal{T}(\psi) - \mathcal{X}_t(\psi))^2- \eta\leq 0.
	\end{equation}
	The ADMM algorithm iteratively updates $\mathcal{X}$, $\{\mathcal{Y}_{k}\}_{k=1}^{n_2}$, $\{\mathcal{Z}_{k}\}_{k=1}^{n_2}$, and $\mu$ until the stop criteria \eqref{stop_criterion_1} and \eqref{stop_criterion_2} are satisfied.
	It follows that $\lVert\mathcal{Z}_{k,t+1}-\mathcal{Z}_{k,t}\rVert_F={\beta_1}\lVert\mathcal{Y}_{k,t+1}-\mathcal{X}_{t+1}\rVert_F\rightarrow 0$, which guarantees the convergence of $\mathcal{Z}_k$.
	The non-positive inequality constraint $\sum_{\psi}\mathcal{W}(\psi)(\mathcal{T}(\psi) - \mathcal{X}_t(\psi))^2 - \eta\leq 0$, and the {projection $(\cdot)^+$} guarantee the convergence of $\mu$.
	HNTC is shown in \textbf{Algorithm} \ref{TC_alg_noisy}.
	

	\begin{algorithm}[ht]
		\caption{Hybrid Noisy Tensor Completion (HNTC)}
		\begin{algorithmic}[1]
			\INPUT incomplete data tensor $\mathcal{T}$, and weighting tensor $\mathcal{W}$ 
			\OUTPUT $\mathcal{T}_c$
			\STATE \textbf{Initialization} 
			$\mathcal{X}_t=\{\mathcal{Y}_{k,t}\}_{k=1}^{n_2}=\mathcal{T}$, $\{\mathcal{Z}_{k,t}\}_{k=1}^{n_2} = {0}$, $\mu_t=0$, $\varepsilon_t=\iota_t=\infty$
			\WHILE{$\varepsilon_t>\epsilon$ {\bf or} $\iota_t>0$}
			\FOR {$\mathbf i_{1:n_2}\in[1,I^\ell_1]\times\cdots\times[1, I^\ell_{n_2}]$}
			\STATE $\mathcal{X}_{t+1}(:,\cdots,:,\mathbf i_{1:n_2})=\mathcal{\hat{X}}$;
			\STATE ($\mathcal{\hat X}$ is obtained by solving \eqref{Noisy_X_unknown_eq})
			\ENDFOR
			\FOR{$\mathbf i_{1: n_1}\in[1,I^s_1]\times\cdots\times[1, I^s_{n_1}]$}
			\STATE  $\mathcal{Y}_{k,t+1}(\mathbf i_{1:n_1},:,\cdots,:)=\mathcal{\tilde{Y}}_k,\ k=1,\dots,n_2$;
			\STATE ($\mathcal{\tilde{Y}}_k$ is obtained by \eqref{Y_sol})
			\ENDFOR
			\STATE Update $\mathcal{Z}_{k,t+1},\ k=1,\dots,n_2$, by 
			\eqref{update_Z};
			\STATE Update $\mu_{t+1}$ by \eqref{update_nu};
			\STATE $\varepsilon_{t+1} =\sum_{k=1}^{n_2}\left\Vert\mathcal{X}_{t+1}-\mathcal{Y}_{k,t+1}\right\Vert_F$; 
			\STATE $\iota_{t+1}=\sum_{\psi}\mathcal{W}(\psi)(\mathcal{T}(\psi) - \mathcal{X}_{t+1}(\psi))^2 - \eta$; 
			\STATE $t:=t+1$;
			\ENDWHILE
			\STATE $\mathcal{T}_c=\mathcal{X}_t$
		\end{algorithmic}
		\label{TC_alg_noisy}
	\end{algorithm}

	\subsection{Online Hybrid Noisy Tensor Completion} \label{Online_TC_algorithm}
	Since the tensor completion problem \eqref{hybrid_tensor_completion_problem} is a very large-scale convex optimization problem, the computing overhead for solving the problem would be quite high, which is challenging in a real-time system.
	For the first channel estimation aided by HNTC, the high computing overhead is not harmful because we can do the offline learning, which means that there is a predefined period of time for building and completing the data tensor before it is applied for the efficient channel training.
	However, for the subsequent channel estimation using HNTC, the data tensor necessitates doing the tensor completion again to provide new predictions exploiting the updated data.
	Thus, the large computing overhead of tensor completion would be impractical in real-time system.

	To address this issue, we introduce the warm-start method \cite{boyd2011distributed}, which initializes the iterative method using the solution obtained from the previous iteration.
	In {HNTC}, we apply an iterative method to solve the completion problem.
	The rate of convergence for this iterative method highly depends on the initial points of the variables, $\mathcal{X},\mathcal{Y}_k,\mathcal{Z}_k$ and $\mu$, in the Lagrange function \eqref{Aug_Lagrange_func}.
	The warm-start method aims to select the initial points of the variables $\mathcal{X},\mathcal{Y}_k,\mathcal{Z}_k, \mu$ close to the convergence point based on the prior knowledge.
	For tensor $\mathcal{X}$, the initial point could be arbitrarily chosen since it is first updated in the iterative algorithm and irrelevant to the initial point of $\mathcal{X}$. 
	For tensors $\mathcal{Y}_k, \mathcal{Z}_k$ and $\mu$, we choose their initial points as $\mathcal{Y}_{k,old}, \mathcal{Z}_{k,old}$ and $\mu_{old}$, which are obtained in the previous iteration.
	Since the data tensor records the average received power which captures the long-term channel conditions, the recorded data are not expected to change drastically for each data update.
	Therefore, the previous ADMM iteration often provides an acceptable guess which leads to fewer iterations than the one with random initialization.
	The performance of the proposed online HNTC is evaluated in Section \ref{Exp_online_learning}.
	\subsection{Complexity Analysis} \label{complexity}
	The computational complexity of HNTC is dominated by the update of $\mathcal{X}$ and $\mathcal{Y}_k$.
	The update of $\mathcal{X}$ requires matrix inversions to solve the linear equations in \eqref{Noisy_X_unknown_eq}, leading to a complexity of $\mathcal{O}\left( \left(\prod_{j=1}^{n_2}I^{\ell}_j\right) \left(\prod_{m=1}^{n_1}I^s_m\right)^3 \right)$.
	For the update of $\mathcal{Y}_k$, the SVD is used to perform the soft-thresholding operation, with the complexity as $\mathcal{O}\left(\left(\prod_{m=1}^{n_1}I^s_m\right) \max_k \left((I_k^{\ell})^2\cdot \prod_{j=1,j\neq k}^{n_2}I^{\ell}_j\right)\right)$ \cite{golub13}.
	


	\section{Numerical Results and Example Applications} \label{sec_numerical_result}
	We evaluate the performance of HNTC with data generated by Quadriga \cite{jaeckel2014quadriga}.		
	We consider an uplink MIMO scenario with carrier frequency $f_c$ using UPAs (as in \eqref{atp}) having $M_r$ antennas at the BS, and $M_t$ antennas at the UE.
	The scenario \textit{mmMAGIC\_UMi\_NLOS} is selected \cite{jaeckel2014quadriga}.
	The simulation parameters are given in Table \ref{table:comm_sim_parameters}.
	The network layout is depicted in Fig. \ref{fig:wireless_comm_scenario}, containing one BS in position $(0,0)$m at height $10$m UEs in the rectangular area $\mathcal G= [10\mathrm{m},60\mathrm{m}]\times [-25\mathrm{m},25\mathrm{m}]$ at height $1.5$m.
	We consider $51\times 51=2601$ reference GPS coordinates uniformly located in the area $\mathcal G$.
	In each of the reference GPS coordinate, we collect the MIMO channel as the ground truth data.
	The position labels are derived as in \eqref{pos_label_derive} within the rectangular region $\bar{\mathcal G}=\mathcal{G}$, with resolution $\Delta_s = 5\ \mathrm{m}$, so that the number of discretized positions in the $x$ and $y$ coordinates are $L_x=L_y=11$.
	Thus, for each position label, we have around $\frac{2601}{11\times 11}\approx 21$ channel measurements.
	
	\begin{table}[h]
		\caption{Common simulation parameters} 
		\label{table:comm_sim_parameters} 
		\centering 
		\begin{tabular}{| r | c|l|} 
			\hline
			Parameter & Symbol & Value\\
			\hline
			Carrier frequency & $f_c$ & $58.68$ GHz\\
			BS antenna number & $M_r$ $(C_y^r,C_z^r)$ & $256$ $(16,16)$\\			
			UE antenna number & $M_t$ $(C_y^t,C_z^t)$ & $16$ $(4,4)$\\
			BS codebook size & $\lvert\mathcal{W}\rvert$ $(C_\theta^r,C_\phi^r)$ & $256$ $(16,16)$\\			
			UE codebook size & $\lvert\mathcal{F}\rvert$ $(C_\theta^t,C_\phi^t)$ & $16$ $(4,4)$\\
			\hline 
		\end{tabular}
	\end{table}	

	The ground truth data representing the noiseless average received power on all combinations of positions and BS codewords is collected in a fourth order tensor $\mathcal{T}_{avg} \in \mathbb{R}^{L_x\times L_y\times C^r_{\theta}\times C^r_{\phi}}$.
	We define the incomplete data tensor $\mathcal{T}$ containing the measured data, along with the tensor $\mathcal V$ denoting the number of measurements collected.
	In our numerical evaluations, we vary the ratio of observed positions $K_{op}=C_{op}/(L_xL_y)$, where $C_{op}$ denotes the number of observed positions.
	Regarding the incomplete data tensor $\mathcal{T}$, we make \textbf{Assumptions \ref{assum_1}}, \textbf{\ref{assum_2}}  for the experiment setting.
	\begin{assumption}\label{assum_1}
		The observed positions $\mathbf{p}$ are randomly chosen from $L_x\times L_y$ grid. For an observed position $\mathbf p'$, the measurements of the reference GPS coordinates corresponding to position $\mathbf p'$, $\{\mathbf g: \rho(\mathbf{g})=\mathbf{p}',\ \mathbf{g}\in \bar{\mathcal G}\}$, are observed.
	\end{assumption}
	\begin{assumption}\label{assum_2}
		For each observed GPS coordinate $\mathbf{g}$, only the measurements of the top $10\%$ beams (ranked by received signal power) are stored in the database.
	\end{assumption}
	With these two assumptions, $\mathcal{T}$ is incomplete in both positions' and beams' dimensions.
	We consider two kinds of collected measurement data: noise-free data and noisy data.
	For noisy data, we assume that channel measurements are collected with a beamforming signal-to-noise ratio at the receiver $\mathrm{SNR}_r=20$ dB.
	Given a MIMO channel $\mathbf{H}$, $\mathrm{SNR}_r$ is defined as 
	\begin{equation}\label{rx_SNR}
		\mathrm{SNR}_r=10\log_{10}\frac{P_t \lVert\mathbf{H}\rVert_2^2}{\sigma_n^2},
	\end{equation}
	where $P_t$ is the transmit power, and $\sigma_n^2$ is the noise variance.

%

	Note that the data of $\mathcal{T}$ may be unavailable in some positions, which means no reconstruction on unknown positions is possible if we only consider the low-rank property.
	To show the advantage of {HNTC}, we compare it with existing tensor completion methods that use both low-rank and smoothness during reconstruction, LRTV-PDS \cite{8625383} and SPCTV \cite{yokota2016smooth}.
	LRTV-PDS considers the low Tucker rank and smoothness during tensor reconstruction; SPCTV is based on the low CP rank prior with a smoothness constraint.
	Since these approaches \cite{8625383,yokota2016smooth} are originally designed for image reconstruction, they consider the data tensor model with low-rank and smoothness consistent through all dimensions, and they assume noiseless measurements in SPCTV or noisy measurements in LRTV-PDS;
	in contrast, HNTC considers smoothness and low rank on the distinct dimensions, and is designed for noisy measurements, by taking into account the number of measurements as weighting contribution to the squared-error term.
	
	\vspace{-1em}
	
	\subsection{Prediction Accuracy Comparison}\label{Exp_01}
\begin{figure}[t!]
		\centering
		\includegraphics[width=.8\linewidth]{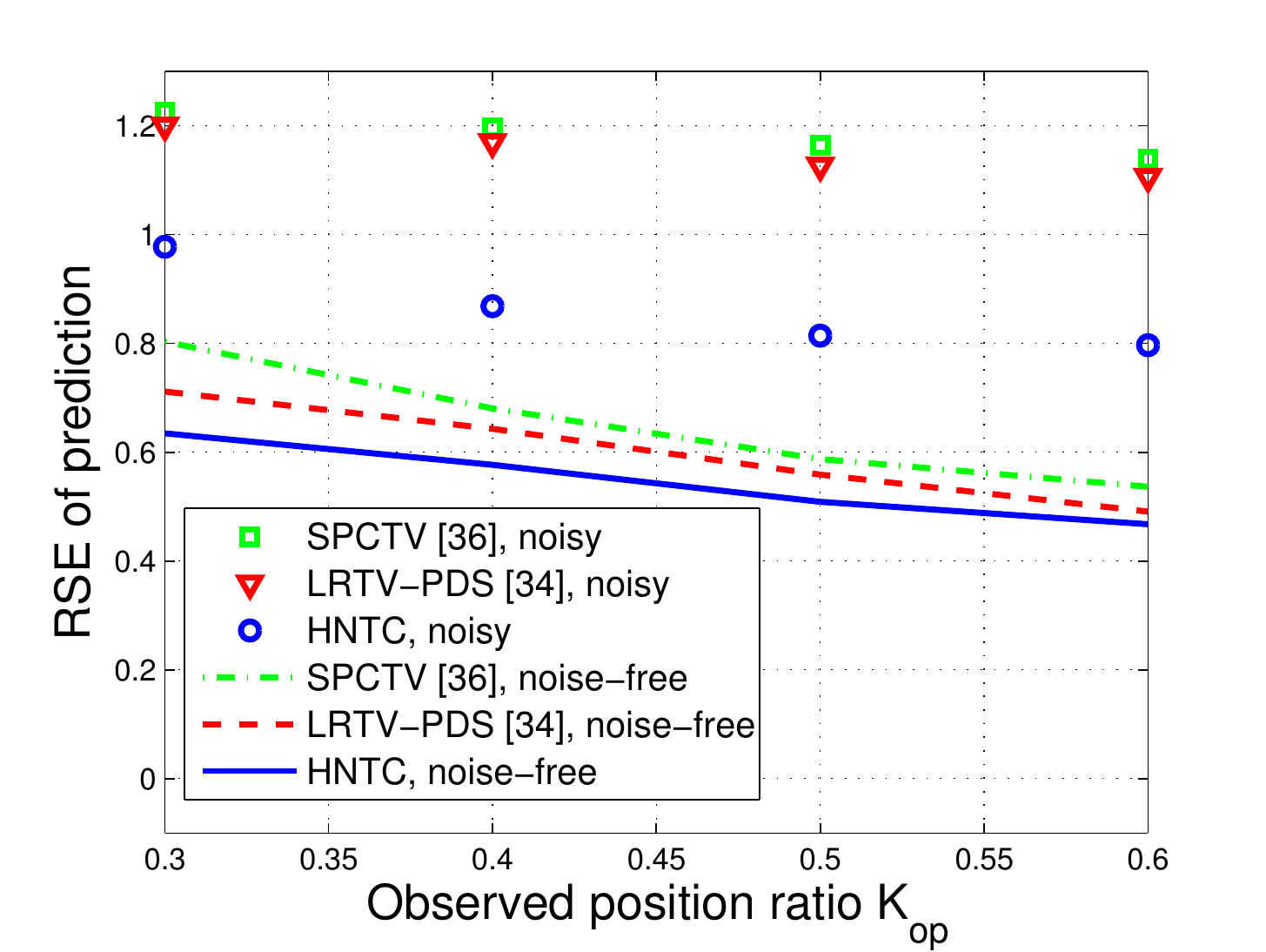}
		\caption{The RSE of prediction versus the ratio of observed positions ($K_{op}$).}
		\label{fig:Exp1}
	\end{figure}

	In Fig. \ref{fig:Exp1}, we evaluate the relative square error (RSE) of the reconstructed tensor versus the observed position ratio $K_{op}$.
	The RSE is defined as 
	\begin{equation}
	\mathrm{RSE}=\frac{\lVert\mathcal{T}_c - \mathcal{T}_{avg}\rVert_F}{\lVert\mathcal{T}_{avg}\rVert_F},
	\end{equation}
	where $\mathcal{T}_c$ is the reconstructed tensor and $\mathcal{T}_{avg}$ is the ground-truth data tensor.
	In Fig. \ref{fig:Exp1}, the trend of RSE is monotonically decreasing with $K_{op}$.
	In fact, with more measurements recorded in the database, the tensor completion algorithms provide better reconstruction.
	We observe that HNTC outperforms both LRTV-PDS and SPCTV in RSE, which means that HNTC provides a better tensor approximation to $\mathcal{T}_{avg}$.
	Given noise-free measurements on only $40\%$ of any possible positions, HNTC attains RSE$=0.57$, as opposed to LRTV-PDS with RSE$=0.64$, and SPCTV with RSE$=0.68$ in the same configuration.
	For the noisy measurement data, the RSE degrades drastically for all approaches.
	Given the noisy measurement tensor on $40\%$ positions, the prediction accuracy attains $\mathrm{RSE}=0.87$ for HNTC, $\mathrm{RSE}=1.17$ for LRTV-PDS, and $\mathrm{RSE}=1.2$ for SPCTV.
	However, HNTC is more robust to noise than the state-of-the-art schemes, since it accounts for noisy measurements as seen in \eqref{sqerr}.



	\vspace{-0.5em}
	
	\subsection{Position-Aided Beam Recommendation} \label{Exp_beam_recomm}
	\subsubsection{Recommendation Algorithm}
	\label{appex}
	For the position-aided beam-recommendation in Section \ref{subsec_position_aided_beam_recomm}, we illustrate the beam subset selection (BSS) algorithm.
	With the completed tensor $\mathcal{T}_c$, we have the estimated 
	received power of all receive beams at UE position $\mathbf{p}$.
	If the number of trained beams is $N_{tr}$, the construction of the 
	recommended beam set is a subset selection problem, that selects the $N_{tr}$ beams with largest estimated received power in the completed tensor $\mathcal{T}_c$, as shown in \textbf{Algorithm} \ref{Beam_subset_selection}.

	\begin{algorithm}[hb!]
		\caption{Beam Subset Selection (BSS)}
		\begin{algorithmic}[1]
			\INPUT completed tensor $\mathcal{T}_c$, number of trained beams $N_{tr}$, BS codebook $\mathcal{W}$ with indices $\mathcal{K}$,			
			UE GPS coordinate $\mathbf{g}$
			\OUTPUT recommended beam subset ${\mathcal S}_{N_{tr}}$
			\STATE \textbf{Initialization} ${\mathcal S}_0\leftarrow\emptyset$
			\FOR{$n=1:N_{tr}$}
			\STATE 
			$(u^*,v^*)=\arg\max_{(u,v)\in
				\mathcal{K}\setminus\mathcal 
				S_{n-1}}\mathcal{T}_c(\rho(\mathbf{g}),u,v)$
			\STATE $\mathcal {S}_n\leftarrow \mathcal {S}_{n-1}\cup 
			{(u^*,v^*)}$
			\ENDFOR
		\end{algorithmic}
		\label{Beam_subset_selection}
	\end{algorithm}


	\subsubsection{Performance of Proposed Beam-Alignment}
	Here, we evaluate the performance of the position-aided beam recommendation with HNTC, compared with the ones with LRTV-PDS and SPCTV.
	Our formulation with tensor completion allows prediction for unknown positions by exploiting spatial correlation.
	However, the state-of-art approach \cite{Va2018} for position-aided beam alignment uses only the prior knowledge available at a given position, and is thus unable to make predictions in positions where the measurements are not available.
	For comparison, we consider the type-B fingerprinting method \cite{Va2018} by providing the recommended beam set based on the closest position having available prior knowledge if the prior measurements of UE position are not given. 
	Moreover, we consider the position-aided beam recommendation aided by BSS with $\mathcal{T}_{avg}$, called \textit{Genie-Aided}.
	The Genie-Aided approach is expected to have the best recommendation performance since it	uses ground truth data at all positions and beams.

\begin{figure}[t!]
	\centering
	\includegraphics[width=.8\linewidth]{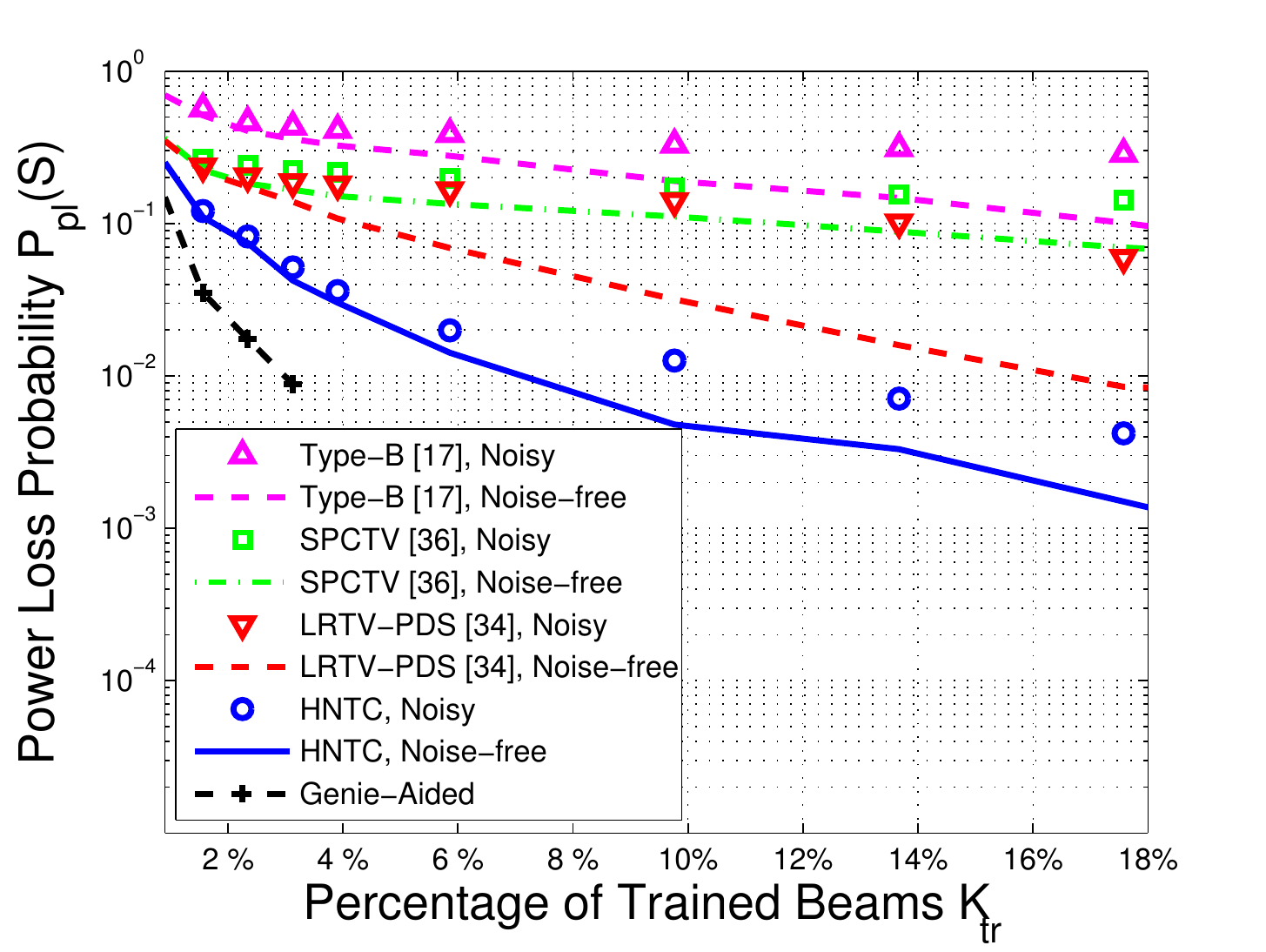}	
	\caption{Power loss probability $(P_{pl}(\mathcal{S}_{\mathbf{p}}))$ versus the percentage of trained beams $(K_{tr})$, with the observed position ratio $K_{op}=40\%$.}
	\label{fig:Exp2_powerloss}
			\vspace{-7mm}
	\end{figure}

	First, we evaluate the power loss probability $P_{pl}(\mathcal{S}_{\mathbf{p}})$ versus the percentage of trained beams $K_{tr}=N_{tr}/\lvert\mathcal{K} \rvert$, where $N_{tr}$ is the number of trained BS beams and $\lvert\mathcal{K} \rvert$ is the total number of BS beams. 
	The set $\mathcal{S}_{\mathbf{p}}$ is the recommended beam subset at position $\mathbf{p}$ using BSS.
	To measure the beam alignment accuracy for the recommended set $\mathcal{S}_{\mathbf{p}}$, we define the power loss probability metric as
	\begin{flalign*}
	P_{pl}(\mathcal{S}_{\mathbf{p}})=\mathbb{P}\Big(\max_{(u,v)\in\mathcal{S}_{\mathbf{p}}}r^{(\mathbf{p},u,v)}<\max_{(u,v)\in\mathcal{K}}r^{(\mathbf{p},u,v)}\Big),
	\end{flalign*}
	where $r^{(\mathbf{p},u,v)}$ is the received power defined in \eqref{eq_rx_pwr_data}.
	The power loss probability is averaged over the channels at the GPS coordinates corresponding to all positions.	
	In Fig. \ref{fig:Exp2_powerloss}, we evaluate the power loss probability versus the percentage of trained beams with the observed position ratio $K_{op}=40\%$.
	For the Genie-Aided approach, it attains $P_{pl}(\mathcal{S}_{\mathbf{p}})=10\%$ with $K_{tr}=1.1\%$.
	With $K_{tr}>3\%$ of  trained beams, the Genie-Aided approach can always recommend the beam set including the best receive beam.
	For the noise-free measurement data tensor, only $1.5\%$ of the trained beams is required for HNTC to attain $P_{pl}(\mathcal{S}_{\mathbf{p}})=10\%$, as opposed to $4\%$ for LRTV-PDS, $12\%$ for SPCTV, and $18\%$ for Type-B.	
	The position-aided beam alignment approach supported by tensor completion (HNTC, LRTV-PDS, or SPCTV) outperforms the state-of-the-art method (Type-B) since the tensor completion provide accurate power prediction on the unavailable positions.
	With the noisy measurement data tensor, the $P_{pl}(\mathcal{S}_{\mathbf{p}})$ supported by our proposed approach (HNTC) is more robust than the ones using LRTV-PDS, SPC-TV, and Type-B. 
	This behavior is in line with the improved performance of HNTC observed in Fig.~\ref{fig:Exp1}.
	
	\subsubsection{Spectral Efficiency}
	Next, we evaluate the spectral efficiency versus the receive SNR.
	We define the achievable transmission rate as 
	\begin{equation} \label{achi_rate}
	R=B\log_2\left(1+ \mathrm{SNR}_{r} \frac{\lVert\mathbf{w}^H \mathbf{H}\mathbf{f}\rVert^2}{\lVert\mathbf{H}\rVert_2^2}\right),
	\end{equation}
	where $B$ is the bandwidth; $\mathrm{SNR}_{r}$ is the receive beamforming SNR defined in \eqref{rx_SNR}; $\mathbf{w}/\mathbf{f}$ is the BS/UE beamforming vector; $\mathbf{H}$ is the MIMO channel.
	The selected BS/UE beamforming vector $(\mathbf{w}^*,\mathbf{f}^*)$ is the best beam pair (ranked by received power) chosen from $\mathcal S\times\mathcal F$, where $\mathcal{S}$ is the BS beam set recommended by BSS, and $\mathcal{F}$ is the UE codebook.
	Since the set of beam pairs $\mathcal S\times\mathcal F$ is scanned exhaustively, the resulting overhead is $T_{train}=(\lvert\mathcal{S}\rvert\cdot\lvert\mathcal{F}\rvert+1)\delta_S$, where the microslot duration $\delta_S=10\ \mu\mathrm{s}$ is the time required to scan a single beam.
	The fraction of time used for data transmission is $f_{comm}=\frac{T_{frame}-T_{train}}{T_{frame}}$, where $T_{frame}=10\ \mathrm{ms}$ is the frame duration.
	The spectral efficiency is defined as $\frac{f_{comm}\times R}{B}$, which accounts for the loss due to the training overhead.
	The proposed recommendation-based method is much more efficient than a conventional exhaustive search method, where all BS and UE beam pairs $\mathcal W\times\mathcal F$ are scanned. In fact, the conventional \textit{exhaustive search} cannot be implemented in our considered scenario since its training overhead exceeds the frame duration, $T_{train}=(\lvert\mathcal{W}\rvert\cdot\lvert\mathcal{F}\rvert+1)\delta_S > T_{frame}$, leaving no time for data transmission.

\begin{figure}[t!]
	\centering
	\includegraphics[width=.8\linewidth]{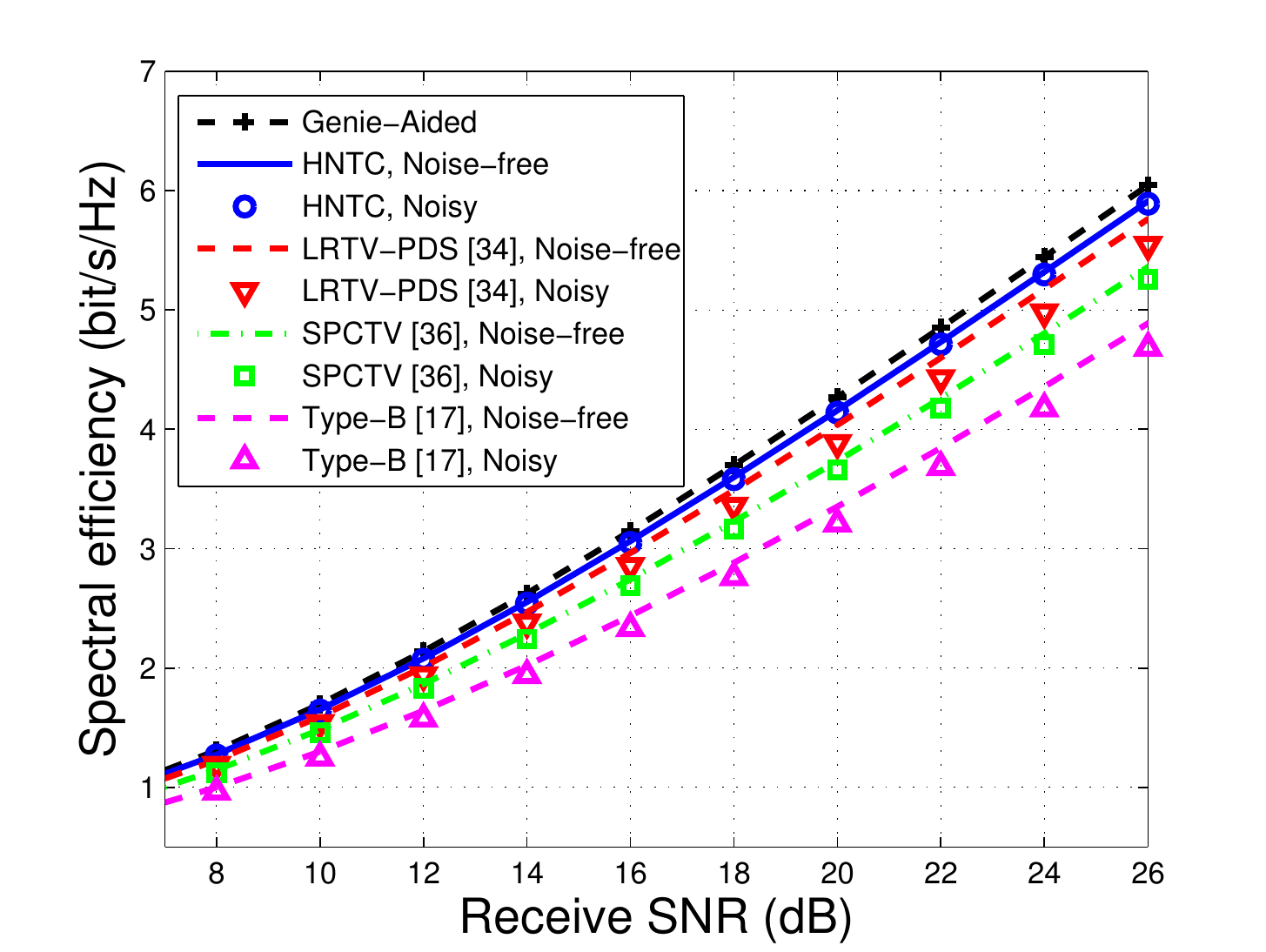}
	\caption{Spectral efficiency $(\frac{f_{comm}\times R}{B})$ versus the receive SNR $(\mathrm{SNR}_{r})$ with $(K_{op},K_{tr})=(40\%,2\%)$.
	}
	\label{fig:Exp2_SpecEffic_Kop40_Ntr5}
	\vspace{-5mm}
\end{figure}

	In Fig. \ref{fig:Exp2_SpecEffic_Kop40_Ntr5}, we evaluate the spectral efficiency $(\frac{f_{comm}\times R}{B})$ versus $\mathrm{SNR}_r$ with $(K_{op},K_{tr})=(40\%,2\%)$.
	The Genie-Aided approach attains the largest spectral efficiency because it has the smallest power loss probability $P_{pl}(\mathcal{S}_{\mathbf{p}})=2.5\%$ with the support of available information on all possible positions.
	For the noise-free measurement data and a reference $\mathrm{SNR}_r=20\ \mathrm{dB}$, the spectral efficiency of HNTC is $4.16$ bit/s/Hz, which is better than the state-of-the-art approach (Type-B) with $3.35$ bit/s/Hz.
	Besides, compared with  the beam recommendation aided by other tensor completion approaches, the spectral efficiency of HNTC outperforms the	one with LRTV-PDS by $0.13$ bits/s/Hz, and the one with SPCTV by $0.43$ bits/s/Hz.
	For the compared approaches, the spectral efficiency with the noisy measurement data is worse than the one with the noise-free measurement data.
	However, our proposed HNTC with noisy measurement data has almost the same spectral efficiency as HNTC with noise-free measurement data.
	HNTC is more robust because $P_{pl}(\mathcal{S}_{\mathbf{p}})$ with noisy measurement data is still fairly low $(<10\%)$ in this configuration (as shown in Fig. \ref{fig:Exp2_powerloss}), which is quite similar to $P_{pl}(\mathcal{S}_{\mathbf{p}})$ with noise-free measurement data.
	

	\subsection{Evaluation of Online Hybrid Noisy Tensor Completion} \label{Exp_online_learning}
	We evaluate the performance of the online HNTC proposed in Section \ref{Online_TC_algorithm} \textbf{with the warm start}, compared with the one \textbf{without warm start}.
	For the prediction accuracy, we observe the RSE between the reconstructed tensor and $\mathcal{T}_{avg}$ and also the number of iterations of ADMM for the convergence rate.
	The noise-free measurement data tensor $\mathcal{T}$ is considered.
	For the online updating scenario, at update instant $0$, we consider the initial ratio of observed positions $K_{op}$ as $K_{ini}=30\%$.
	We assume that the measurements of $N_{upd}=5$ new positions are updated to the data tensor $\mathcal{T}$ in each subsequent update instant.
	For HNTC \textbf{with warm start}, we initialize the variables in ADMM as the ones retrieved from the previous update instant.
	
\begin{figure}[t]
		\centering
		\includegraphics[width=.8\linewidth]{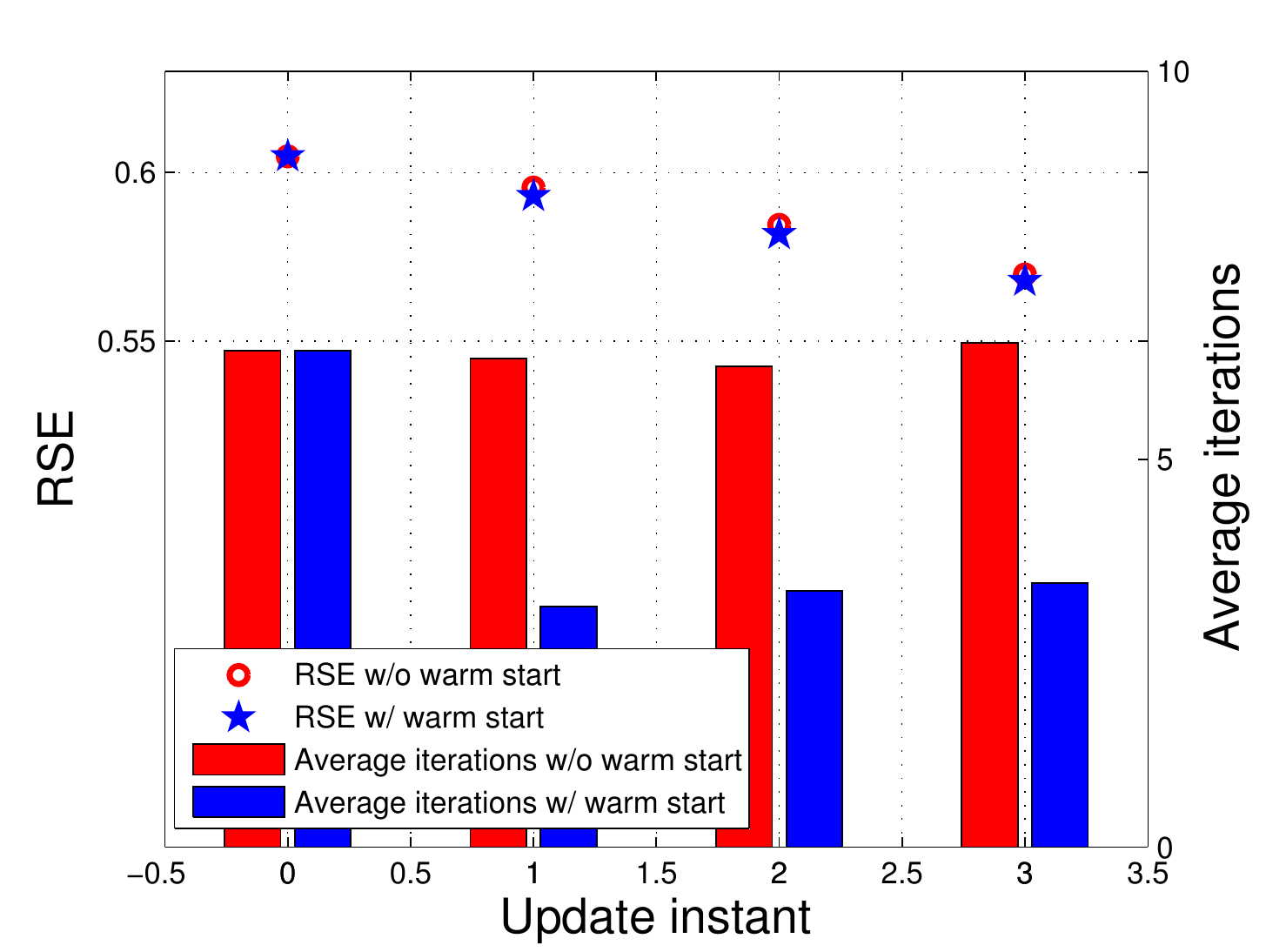}		
		\caption{RSE and average number of iterations comparison between {HNTC} \textbf{with warm start} and {HNTC} \textbf{without warm start}, $(K_{ini}, N_{upd})=(30\%,5)$.}
		\label{fig:Exp4_OnlineTC}
			\vspace{-3mm}
	\end{figure}

	In Fig. \ref{fig:Exp4_OnlineTC}, for the RSE comparison, the HNTC \textbf{with warm start} is close to the HNTC \textbf{without warm start}.
	In update instant $0$, the average number of iterations of both cases is around $6.3$ because there is no available prior information.
	For the subsequent update instants, HNTC \textbf{with warm start} converges in around $3.2$ iterations, while HNTC \textbf{without warm start} requires $6.3$ iterations.
	The warm start method reduces the computational complexity by converging in $50\%$ fewer iterations without compromising on the prediction accuracy.

	\vspace{-2mm}
	\subsection{Noisy Positional Information} \label{Exp_time_varying_channel}
	Here, we investigate the influence of noisy positional information on the position-aided beam recommendation.
	In a practical setting, the positional information is acquired via the process of GNSS/GPS estimation \cite{roy2014smartphone,maschietti2017robust}.
	Due to the mobility and estimation error, the obtained positional information may be noisy, which would impair the performance of the position-aided channel estimation.
	To model the positional error, let $E(d)=\{\mathbf{x}\in\mathbb{R}^2: \lVert\mathbf{x}\rVert_2 \leq d\}$ be a two dimensional closed disk centered at the origin with radius $d$; we then model the random error as $\mathbf{e}\in\mathbb{R}^2$ uniformly distributed in $E(d)$  \cite{maschietti2017robust}.
	The noisy spatial coordinate is defined as $\mathbf{g}_r = \mathbf{g}+\mathbf{e}$, where $\mathbf{g}$ is the ground truth.
	
	To alleviate the impairment of the noisy spatial coordinates, we propose the grouping-based beam subset selection {(G-BSS)} in \textbf{Algorithm} \ref{Group_Beam_subset_selection}.
	Given the received noisy spatial coordinate $\mathbf{g}_r$ and error radius $d$, we collect all possible positions as a set $\mathcal{P}=\{\rho(\mathbf{g}):\lVert\mathbf{g}-\mathbf{g}_r\rVert_2\leq \zeta d\}$, where $\zeta$ is a constant coefficient. 
	Note that the performance is influenced by the selection of $\zeta$, and we choose $\zeta=0.4$ in this work. 
	Then, we derive the subtensor $\bar{\mathcal{R}}=\frac{1}{\lvert\mathcal{P}\rvert}\sum_{\mathbf{p}\in\mathcal{P}}\mathcal{T}_c(\mathbf{p},:,:),$ which contains the predicted received power of each beam by averaging over all possible positions in the set $\mathcal{P}$.
	Finally, we use the tensor $\bar{\mathcal{R}}$ to provide the beam recommendations for the UE.

\begin{figure}[t!]
		\centering
		\includegraphics[width=.8\linewidth]{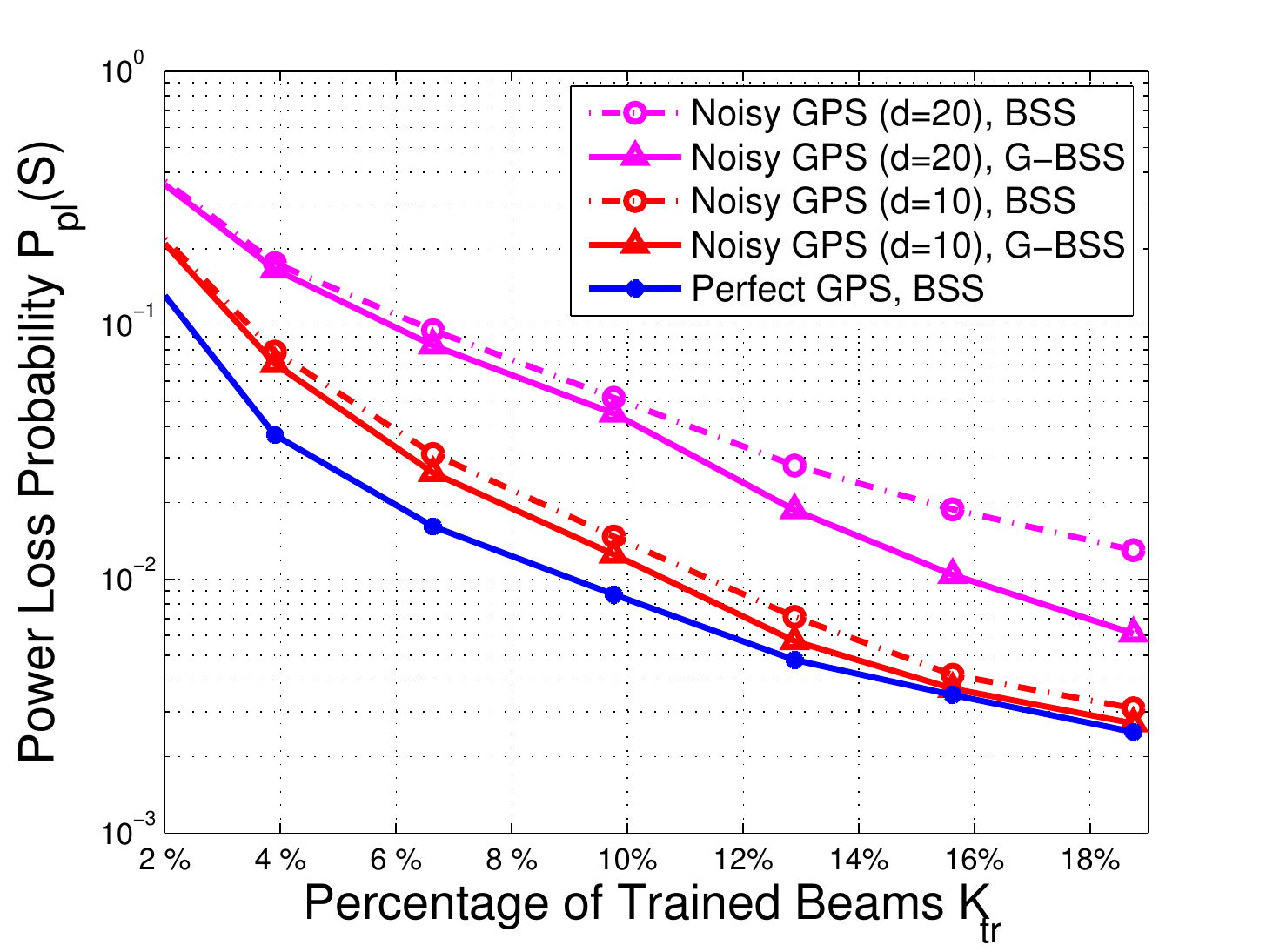}
		\caption{Power loss probability versus the percentage of trained beams in the scenario with noisy positional information.}
		\label{fig:Exp5_pos_error}
	\end{figure}

	In Fig. \ref{fig:Exp5_pos_error}, we evaluate the power loss probability versus the percentage of trained beams, under the scenario with noisy positional information.
	We compare the performance of G-BSS with that of BSS, which neglects the error in the positional information.
	The completed tensor $\mathcal{T}_c$ is reconstructed by {HNTC}, and the noise-free measurement data is considered.
	With the BSS in \textbf{Algorithm} \ref{Beam_subset_selection}, the power loss probability increases when the positional error radius $d$ increases.
	The performance of the position-aided beam recommendation is impaired by the noisy positional information.
	However, G-BSS is more robust than BSS against these impairments.
	Given $K_{tr}=10\%$, the BSS with perfect GPS attains $P_{pl}(\mathcal{S}_{\mathbf{p}})=0.87\%$.
	For the scenario with positional error $d=10\ \mathrm{m}$, the power loss probability is improved by G-BSS to be $P_{pl}(\mathcal{S}_{\mathbf{p}})=1.2\%$, compared with $P_{pl}(\mathcal{S}_{\mathbf{p}})=1.5\%$ by BSS.
	If the positional error increases to $d=20\ \mathrm{m}$ , the power loss probability is improved with the support of G-BSS to be $P_{pl}(\mathcal{S}_{\mathbf{p}})=4.5\%$, as opposed to $P_{pl}(\mathcal{S}_{\mathbf{p}})=5\%$ by BSS.
	
	\begin{algorithm}[h!]
		\caption{Grouping-based Beam Subset Selection (G-BSS)}
		\begin{algorithmic}[1]
			\INPUT completed tensor $\mathcal{T}_c$, number of trained beams $N_{tr}$, BS codebook $\mathcal{W}$ with indices $\mathcal{K}$,
			UE noisy GPS $\mathbf{g}_r$, positional error radius $d$
			\OUTPUT recommended beam subset ${\mathcal S}_{N_{tr}}$
			\STATE \textbf{Initialization} ${\mathcal S}_0\leftarrow\emptyset$
			\STATE $\mathcal{P}=\{\rho(\mathbf{g}):\lVert\mathbf{g}-\mathbf{g}_r\rVert_2\leq \zeta d\}$
			\STATE $\bar{\mathcal{R}}=\frac{1}{\lvert\mathcal{P}\rvert}\sum_{\mathbf{p}\in\mathcal{P}}\mathcal{T}_c(\mathbf{p},:,:)$
			\FOR{$n=1:N_{tr}$}
			\STATE 
			$(u^*,v^*)=\arg\max_{(u,v)\in
				\mathcal{K}\setminus\mathcal 
				S_{n-1}}\bar{\mathcal{R}}(u,v)$
			\STATE $\mathcal {S}_n\leftarrow \mathcal {S}_{n-1}\cup 
			{(u^*,v^*)}$
			\ENDFOR
		\end{algorithmic}
		\label{Group_Beam_subset_selection}
	\end{algorithm}

	\vspace{-2mm}
	\section{Conclusions} \label{sec_conclusion}
	In this paper, we proposed a learning framework to perform data-assisted beamforming in MIMO communication over a fixed service area with noisy power measurements on a small subset of possible positions.
	In our model, the received power and side information (e.g., user positions and receive beams) were collected into a data tensor.
	We developed a noisy tensor completion, HNTC, exploiting the low-rank and smoothness properties of the channel data.
	The numerical results showed that HNTC provides more accurate received power prediction than the state-of-the-art tensor completion method \cite{8625383,yokota2016smooth} utilizing both the low-rank and the smoothness of the data.
	Furthermore, the beam recommendation aided by HNTC was shown to improve the performance of beam alignment over the state-of-the-art data-assisted beam alignment approach \cite{Va2018}, by improving the prediction accuracy and reducing the beam training overhead.
	


	
	
	\bibliographystyle{IEEEtran}
	\bibliography{IEEEabrv,reference}

\end{document}